\documentclass[final,5p,times,twocolumn,authoryear,fleqn]{elsarticle}
\pdfoutput=1

\usepackage[utf8]{inputenc}

\usepackage{amssymb}
\usepackage{amsthm}
\usepackage{amsmath}
\usepackage{pdflscape}
\usepackage{comment}
\usepackage{longtable}
\usepackage{booktabs}
\usepackage{soul}
\usepackage[normalem]{ulem}

\newcommand{\ut}{{\mathbf{u}}} 
\newcommand{\ph}{\phi} 

\newcommand{\up}{u_{\ph}}

\usepackage[colorlinks=True,%
            urlcolor=blue,%
            linkcolor=blue,%
            citecolor= blue]{hyperref}

\usepackage{lineno}

\journal{Icarus}

\begin{document}

\begin{frontmatter}


\title{Zonostrophic turbulence in the subsurface oceans of the Jovian and Saturnian moons}

\author[inst1]{Simon Cabanes}
\author[inst1]{Thomas Gastine}
\author[inst1]{Alexandre Fournier}

\affiliation[inst1]{%
organization={Université Paris Cité, Institut de physique du globe de Paris, UMR 7154 CNRS}, 
addressline={1 rue Jussieu}, 
postcode={F-75005},
city={Paris}, 
country={France}}
            
\begin{abstract}
In order to characterize the global circulation of the subsurface ocean of Jovian and Saturnian moons, we analyze the properties of 21 three-dimensional simulations of Boussinesq thermal convection in a rapidly rotating spherical shell. Flow is driven by an adverse temperature contrast imposed across the domain, and is subjected to no-slip boundary conditions. We cover a region of parameter space previously unexplored by global simulations, both in terms of rapid rotation and vigor of convective forcing, closer to, yet still admittedly far from, the conditions appropriate for the subsurface ocean of Ganymede, Europa, Enceladus, and Titan. Our most extreme simulations exhibit a dynamic global circulation that combines powerful east-west zonal jets, planetary waves, and vortices. A spectral analysis of the kinetic energy distribution performed in cylindrical geometry reveals a high degree of anisotropy of the simulated flows. Specifically, the axisymmetric zonal energy spectra follow a steep $-5$ slope in wavenumber space, with the energy amplitude exclusively controlled by the rotation rate. In contrast, the non-axisymmetric residual spectra display a gentle $-5/3$ slope, with the energy amplitude controlled by the thermal buoyancy input power. This spectral behavior conforms with the theory of zonostrophic turbulence, as coined by \cite{sukoriansky02}, and allows us to propose tentative extrapolations of these findings to the more extreme conditions of icy satellites.
By assuming that kinetic energy dissipates via Ekman friction at the ice-ocean boundary, we predict an upper 
bound for the geostrophic zonal velocity ranging from a few centimeters per second for 
Enceladus to about one meter per second for Ganymede, with residual velocities 
smaller than the zonal velocity by an order of magnitude on each moon. These predictions yield typical jets size approaching the ocean depth of Titan, Ganymede and Europa and  $10$ to $40\%$ of the ocean depth on Enceladus.
\end{abstract}




\end{frontmatter}


\section{Introduction}

The astrobiological potential of the Jovian and Saturnian moons, 
most notably Europa, Ganymede, Enceladus, and Titan, has come to the
fore in the wake of the Galileo, Cassini-Huygens, and Juno space missions. 
These satellites contain more than water: 
non-aqueous components, including salts and sulfur, have been detected on 
the surface of Enceladus, while organic materials have been identified within plumes emanating from geysers at its South pole \citep{postberg11,postberg18}. 
While \cite{hsu15} have provided persuasive indications of silica nanoparticles 
on the same Enceladus, 
  spectroscopic analyses have revealed the presence of diverse salts, such as MgSO$_4$, NaCl, NH$_4$Cl, as well as sulfur compounds and organic materials on
the surfaces of Europa \citep{carlson09} and Ganymede \citep{tosi23}. This catalog
of  observations points to a well-mixed subsurface ocean and a substantial
 hydrothermal activity inside these moons. 
 
 The concealed ocean of these icy satellites plays a crucial role in shaping their surface
composition \citep[see the review by][]{soderlund20}. A good description 
of its circulation is key, since it governs the transport of heat and matter
 from the floor up to the icy crust and back.
 To date, most of our knowledge on the dynamics of subsurface  oceans relies on the outcome of global first-principle simulations. These simulations drive oceanic flow either by thermal convection \citep[e.g.][]{soderlund14}, mechanical 
 forcing \citep[e.g., tides or libration][]{grannan17,lemasquerier17}, or even magnetic forcing due to electromagnetic pumping in the case of Europa \citep{gissinger19}. 
Recent numerical investigations have also explored the effects of heat flux variations from the inner mantle \citep{terra23} and temperature fluctuations beneath the ice shell on oceanic convection dynamics \citep{kang22}.  
In these studies, the resulting flows promote turbulent mixing within the bulk ocean and thermo-compositional gradients, which have significant implications for the ice shell, transport properties, surface geology, and potential habitability. 
Nevertheless, bridging the gap between this research and \textit{in situ} observations remains an intricate challenge, demanding further exploration and validation.

Addressing this challenge, several studies have proposed the possibility of connecting
thermal convection in the subsurface ocean to the variations of the ice shell 
thickness inferred from space missions \citep{kvorka18}. Numerical simulations of rotating convection in spherical shells conducted by \cite{amit20}, \cite{soderlund14,soderlund19}, \cite{kvorka22} and \cite{gastine23} demonstrate that planetary rotation results in a pronounced latitudinal dependence of the flow, which impacts the heat flux across the outer surface. Regions with high heat flux are more likely to experience increased melting, providing a potential link 
between observed ice shell topography and the behavior of 
the underlying ocean. However, no consensus has emerged 
among these studies regarding the flow patterns 
characterizing each satellite.
This lack of consensus may stem from variations in the boundary conditions prescribed by the authors, as well as 
from the discrepancies between the planetary regimes and those achieved numerically. 

The forthcoming NASA's Europa Clipper mission \citep{roberts23} and ESA's JUICE mission \citep{grasset13} will provide additional data on surfacial heat flux and ice topography, 
which should reinvigorate ongoing debates and shed new light on ocean dynamics.

In the meantime, we propose in this study to resort to a predictive theoretical approach to unveil the dynamics of the subsurface oceans, capitalizing
 on progress made over the past two decades
  in the topic of rotating turbulence. 
 Our working assumption is that oceanic flow is driven by thermal convection alone. We will calibrate our theory using 
 a  suite of numerical simulations, for subsequent 
  extrapolation to the oceans of the Jovian and Saturnian moons. 

The simulations involve a spherical fluid layer of thickness $D$ confined between 
two rigid boundaries, which rotates at a constant rotation rate $\Omega$. 
Convective motions are driven by a fixed temperature contrast $\Delta T$ imposed
between the two boundaries. The ratio of the inner radius of the shell to its outer radius is set to 0.8. 
The fluid mechanical problem at hand 
is controlled by three dimensionless numbers: the Ekman number $E = \nu/\Omega D^2$, the Rayleigh number $Ra = \alpha g_o D^3 \Delta T/ \nu \kappa$, and the Prandtl number $Pr = \nu/\kappa$, where $\nu$ and $\kappa$ are the viscous and thermal diffusivities, $\alpha$ is the thermal expansivity, and $g_o$ is the acceleration of gravity at the surface of the domain. 
\begin{figure*}[h!]
  \centering
  \includegraphics[width=.8\textwidth]
    {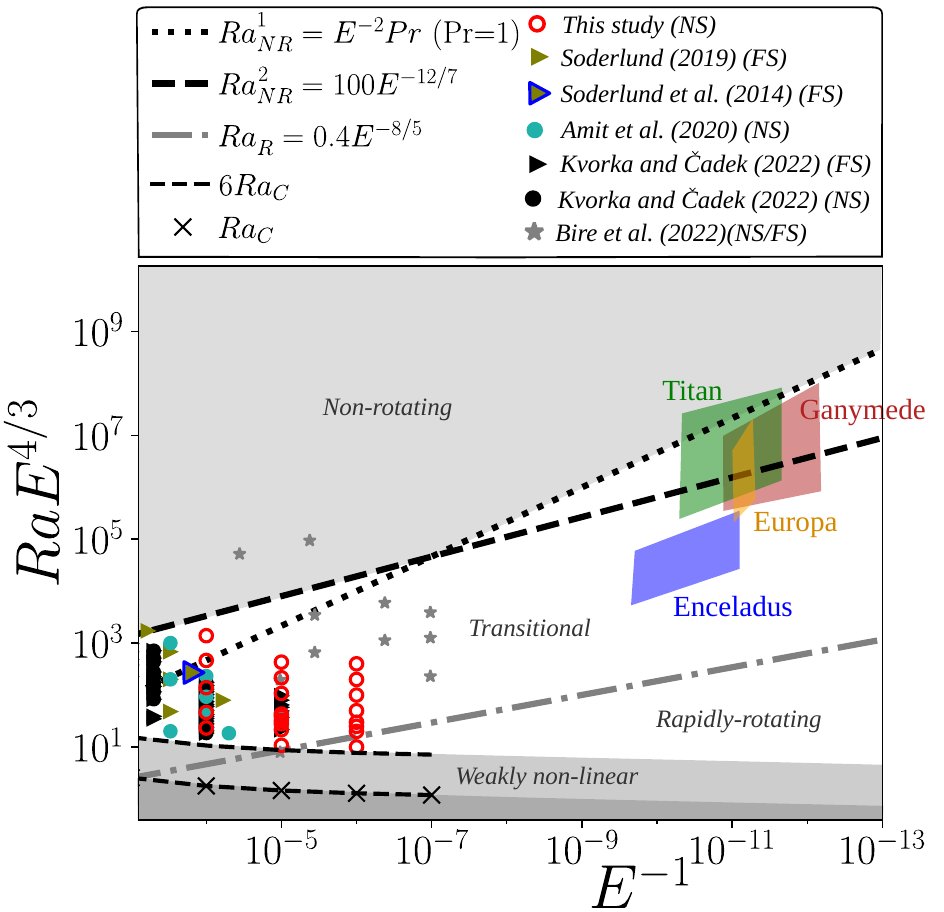}
    \caption{Convective regime diagram following \cite{gastine16} with superimposed parameter estimates for Enceladus (blue), Titan (green), Europa (yellow), and Ganymede (red) as outlined in \ref{Ap:param} and Table~\ref{tab:Annexe1} and following the calculation provided in \cite{soderlund19}.
    Direct numerical simulations from the current study and from \cite{soderlund19,amit20,kvorka22} and \cite{bire22} are reported with circles/triangles for no-slip(NS)/free-slip(FS) boundary conditions.
    The black crosses denote the critical Rayleigh number $Ra_C$ given by \cite{barik23} for each Ekman number and a radius ratio of $0.8$. The boundary $6\,Ra_C$ delimits the weakly non-linear regime defined by \cite{gastine16} for a radius ratio of $0.6$. 
    Straight lines are predicted transitions from the regime of geostrophic turbulence to a transitional regime less influenced by rotation (grey line), and from the transitional regime to a non-rotating regime (black lines). Scaling laws
    for $Ra_R$ and $Ra_{NR}^1$ are taken from \cite{gastine16} and 
    for $Ra_{NR}^2$ from \cite{gilman77}. 
    } \label{fig:RavsEk}
\end{figure*}

In this context, it is common practice to refer to the regime diagram for rotating convection initially introduced by \cite{gastine16}. 
Figure~\ref{fig:RavsEk} replicates and extends this diagram by incorporating additional studies that deal with the subsurface oceans of the icy moons. 
The onset of convection is determined by the critical Rayleigh number $Ra_C$ and is expected to follow $Ra_C \sim E^{-4/3}$ in the limit of vanishing Ekman numbers. The exact values shown
by black crosses in Fig.~\ref{fig:RavsEk} come from the linear computations by \cite{barik23} for a spherical shell with our adopted radius ratio of $0.8$. 
The parameter space $(E,RaE^{4/3})$ is divided into several dynamical regimes. 
Beyond the weakly non-linear regime close to onset, and for Rayleigh numbers below $Ra_{R}$, the influence of rotation is maximum,  resulting in a predominantly geostrophic flow where the balance between Coriolis force and pressure gradient leads to a strong flow invariance along the direction of rotation.
For Rayleigh numbers exceeding $Ra_{NR}$, the influence of rotation is lost and the flow is dominated by fluid inertia. 
In between, a transitional regime prevails, with a weak but still significant influence of rotation.
Given the ongoing debates regarding the scaling laws for $Ra_{NR}$, we report in Fig.~\ref{fig:RavsEk}
two possible transition parameters \citep[see ][ for more details]{gastine16, cheng18}. Note also that these regime boundaries are susceptible to change with the radius ratio of the fluid domain. 

In addition to the new calculations performed for this study, Fig.~\ref{fig:RavsEk} also includes simulations of 3D convection by 
\cite{soderlund14, amit20, kvorka22} and \cite{bire22}.  
The numerical cost of such global simulations makes it necessary to operate with enhanced values of the 
fluid kinematic viscosity $\nu$ and thermal diffusivity $\kappa$. Employing 
physically sound no-slip (also termed rigid) boundary conditions, 
as done e.g. by \cite{amit20}, 
may tend to exaggerate the role played by viscous stresses in the dynamics. 
This led several authors \citep[e.g.][]{soderlund14,soderlund19,kvorka22} to opt for the more questionable free-slip boundary 
conditions. 
Additionally, simulations from \cite{bire22}, computed in a spherical  wedge geometry, considered a hybrid combination of no-slip at the bottom boundary and free-slip at the top boundary. 

The generic picture that emerges from these previous studies is that convective motions are significantly influenced by the relative importance of rotation. Rotation enforces flow invariance along the rotation axis and fosters the formation of large-scale, axisymmetric (invariant in azimuth) flows known as zonal jets \citep[e.g.][]{christensen02}.
\cite{bire22} show that the zonal flow pattern takes the form of  multiple jets of alternated directions with a weak prograde jet at the equator (see their Fig.~6). 
At larger convective forcings (when $Ra\sim E^{-2}Pr$), the zonal flow pattern 
  transitions to a three-jet configuration dominated by a strong retrograde equatorial jet \citep[see, e.g.][her Fig.~2]{soderlund19}.
In these prior studies, free-slip boundary conditions have been shown to promote stronger jets than those obtained in a no-slip configuration. 
When free-slip boundaries are employed, zonal flows dominate the kinetic energy budget, for instance reaching up to 98\% of the total energy at $E=10^{-5}$ in the simulations by \cite{yadav16}, while this fraction is much smaller in simulations that adopt no-slip boundaries, for instance $\sim$ 50\% at $E=3\times 10^{-7}$ in \cite{gastine23}.

Following \cite{soderlund19}, one can estimate the Ekman and Rayleigh numbers of the icy satellites and locate them in the regime diagram (see the coloured polygons in Fig.~\ref{fig:RavsEk}).
Since these estimates fall within the transitional regime, the flow of the subsurface oceans is expected to be influenced by planetary rotation.
Also, the circulation of Enceladus is likely to be more rotationnaly-constrained than that of Ganymede, Europa, and Titan, which are closer to the boundary of the non-rotating regime.  

In any event, the force balance of interest in this study is such that 
the Coriolis force dominates both viscous stresses and, to a lesser extent, inertia. 
In terms of dimensionless numbers, this implies that $E\ll 1$ and that  the Rossby number $Ro = U/\Omega D$ (here $U$ is a typical flow velocity) is below unity. 
Such flows, because they are essentially two-dimensional (2D), transfer energy upscale through non-linear processes known as turbulent cascades \citep{Kraichnan67}. 
In addition, the spherical geometry of planetary fluid layers enforce a zonal anisotropization of the turbulent cascades to form east-west jets via the so-called $\beta$-effect \citep{rhines75}.  These specific planetary conditions define the regime of so-called zonostrophic turbulence \citep{sukoriansky02}.

\begin{figure}[h!]
  \centering
  \includegraphics[width=.49\textwidth]
    {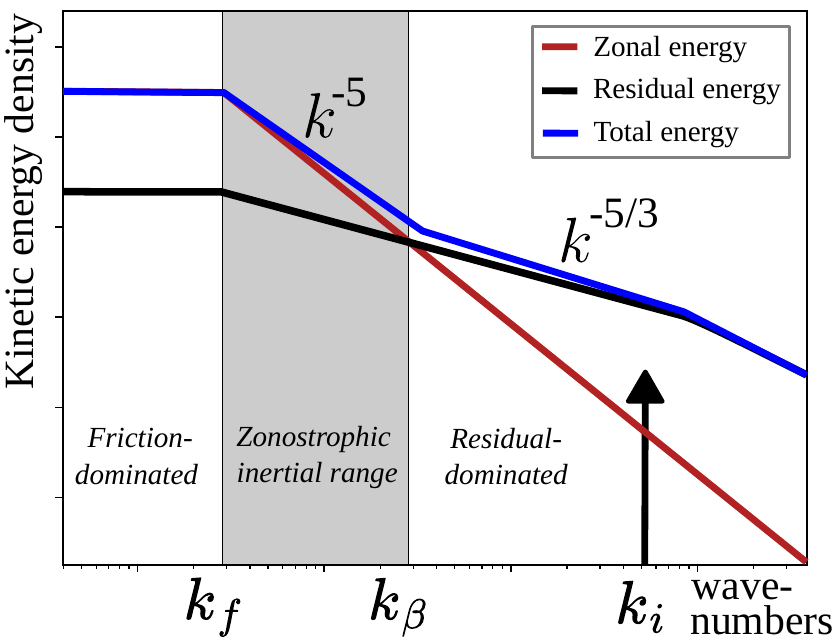}
    \caption{Theoretical kinetic energy distribution as a function of wavenumber $k$ in zonostrophic turbulence, after \cite{galperin10} and \cite{cabanes20a}.}
    \label{fig:zonostrophy}
\end{figure}

In the framework of 2D-turbulence, \cite{sukoriansky02} provided scaling laws that govern the statistical distribution of kinetic energy in zonostrophic flows. 
Figure~\ref{fig:zonostrophy} illustrates this theoretical distribution against wavenumber $k$, where each wavenumber corresponds to a specific scale of motion. 
Zonostrophic turbulent flows exhibit universal kinetic energy spectra. The zonal (axisymmetric) component follows a steep $k^{-5}$ slope, 
while the residual (non-axisymmetric) component adheres to the classical
Kolmogorov-Batchelor-Kraichnan (KBK) scaling with a $k^{-5/3}$ slope. 
Also, the magnitude of the zonal spectrum solely depends on the rotation rate and some geometric parameters of the fluid shell, while the magnitude of 
the residual spectrum is determined by the turbulent power injected at wavenumber $k_i$ (see below for details). 
According to \cite{sukoriansky02}, a flow is considered zonostrophic when an inertial range exists between the friction-dominated wavenumber $k_f$ and the residual-dominated wavenumber $k_{\beta}$, as illustrated in Fig.~\ref{fig:zonostrophy}.
In that range, zonal energy surpasses its residual counterpart. Large-scale Ekman friction, which is contingent upon
the prescription of no-slip boundary conditions,
impedes energy transfer to smaller wavenumbers ($k<k_f$). 
This upscale transfer is possible for free-slip simulations, which may lead to overestimate the global kinetic energy budget.

To date, spectral distributions such as the one sketched  in Fig.~\ref{fig:zonostrophy} have been detected in Jupiter’s tropospheric winds \citep{galperin14,young17}, in mechanically forced laboratory experiments \citep{read15,cabanes17,galperin16,lemasquerier23}, in numerical simulations 
on a rotating sphere \citep{huang01,galperin06} as well as in global circulation atmospheric models \citep{cabanes20a}. 
However, this particular regime has not yet been observed in analog experiments replicating rotating convection.

Indeed, despite the substantial body of research demonstrating that rotating convection exhibits the essential ingredients of zonostrophic turbulence, 
it remains a loosely-documented aspect. For instance, upscale turbulent cascades 
have been studied in three-dimensional (3D) simulations of rapidly rotating Rayleigh-B\'enard convection \citep{favier14,kunnen16,maffei21}. 
Carrying out a flow analysis, \cite{boning23} provided compelling evidence that planetary jets can be driven through statistical correlations of small turbulent scales in 3D spherical convection.
Additionally, the formation of zonal jets has been observed in simulations replicating conditions akin to those found within the Earth's liquid core \citep{guervilly17} and the deep convective envelope of Jupiter and Saturn \citep{aurnou01,heimpel05,yadav20}. In light of these findings, our first goal is to demonstrate the adequacy of zonostrophic theory to account for the statistical properties of rotating convective flows located in the transitional region of Fig.~\ref{fig:RavsEk}, keeping in mind for later that the applicability of this theory might become questionable when approaching the regime of non-rotating convection.




Subsequently, we intend to make use of this theoretical framework to assess the kinetic energy budget and flow regimes within the oceans of Europa, Ganymede, Enceladus, and Titan. Previous attempts to predict subsurface ocean flows have relied on scaling arguments, drawing an hypothetical equivalence between the Rossby number of the numerical simulations and that of the actual oceanic flows \citep[e.g.][]{vance21,bire22}. In addition, since the simulations in question adopt free-slip boundary conditions, they neglect friction effects, which may lead to overestimate the global energy budget.
Alternatively, \cite{jansen23} utilize energetic constraints to infer flow velocity within the subsurface oceans, under the assumption that 
dissipation is controlled by a turbulent quadratic drag akin to that of the Earth's 
ocean \citep{jansen16}. Employing a comparable approach, but with a different assumption regarding the nature of dissipative processes,
we provide an upper bound for the zonal flow velocity. 
Drawing upon the theory of zonostrophic turbulence, we 
derive the typical scale of zonal 
jets in subsurface oceans, along with the residual velocity.

To this end, we conduct $21$ high-resolution simulations of 3D convective turbulence in a spherical shell. 
We explore a parameter range that reaches Ekman numbers smaller by at least one order of magnitude
with respect to previous studies (recall~Fig.~\ref{fig:RavsEk}).  
We present in Section~\ref{sec:Hydrodynamical model} the governing equations, numerical method, 
parameter coverage, and key diagnostics used in this study. 
In Section~\ref{sec:Numerical results}, we present global properties of the kinetic energy in our numerical simulations and we conduct a statistical analysis of the zonal (axisymmetric) and residual (non-axisymmetric) flow components using cylindrical harmonic functions.
In Section~\ref{sec:Implications for subsurface oceans} we delve into the insights offered by the theory of zonostrophic turbulence to explore the icy satellites' oceans. Thanks to estimates of the oceans' depth, the rotation rate of the moons and the heat flux coming up from their interiors we produce predictive kinetic energy budget of Enceladus, Titan, Europa and Ganymede's oceans. 
Finally, in Section~\ref{sec:Discussion}, we introduce a novel 
 and complementary  regime diagram that combines 
  Ekman and Rossby numbers.

\section{Hydrodynamical model}\label{sec:Hydrodynamical model}
\subsection{Governing equations}
We consider rotating convection of a Boussinesq fluid confined in a spherical shell. The fluid shell rotates at a constant angular frequency $\Omega$ about the axis $z$.
Convective motions are driven by a fixed temperature contrast $\Delta T = T_i - T_o$ between the inner radius $r_i$ and the outer radius $r_o$ of the sphere.
Boundaries are impermeable, no-slip and held at constant temperatures. We adopt a dimensionless formulation of the Navier–Stokes equations using the shell 
thickness $D= r_o - r_i$ as the reference length scale and 
the inverse rotation rate $\Omega^{-1}$ as the reference time scale.
The temperature contrast $\Delta T$ defines the temperature scale and gravity is non-dimensionalised using its reference value at the outer boundary $g_o$.
The dimensionless equations that govern convective motions for the velocity $\ut$, the pressure $p$ and the temperature $T$ are expressed by
\begin{equation}\label{eq:governing}
\begin{split}
    \nabla \cdot {\mathbf{u}} &= 0,\\
    \frac{\partial \ut}{\partial t} + \ut \mathbf{\cdot \nabla} \ut + 2 \mathbf{e}_z \times \ut &= -\nabla {p} + \frac{Ra\,E^2}{Pr} {g} {T} \mathbf{e}_r + E\mathbf{\nabla}^2 \ut,\\
    \frac{\partial {T}}{\partial t} + \ut \mathbf{\cdot} \mathbf{\nabla} {T}  &= \frac{E}{Pr} \nabla^2 {T}\,.
\end{split}
\end{equation}

The unit vectors in the radial and vertical directions are denoted by $\mathbf{e}_r$ and $\mathbf{e}_z$, respectively. 
The system of equations~\eqref{eq:governing} is governed by three dimensionless numbers, the Ekman number $E$, the Rayleigh number $Ra$ and the Prandtl number $Pr$ defined above. The radius ratio of the spherical shell is $\eta = r_i/r_o$. To compare with previous studies, we adopt a linearly varying  gravity with $g=r/r_o$. 

\subsection{Numerical method and parameter coverage}
Numerical simulations have been computed using the open source pseudospectral code \texttt{MagIC}\footnote{available at \url{https://github.com/magic-sph/magic}} \citep[the reader is referred to][for more details]{wicht02,gastine16} and the open-source library \texttt{SHTns} for the spherical harmonic transforms \citep{schaeffer13}\footnote{available at \url{https://gricad-gitlab.univ-grenoble-alpes.fr/schaeffn/shtns}}.
The system of equations~\eqref{eq:governing} is solved in spherical coordinates $(r, \theta, \phi)$ and the velocity and temperature fields are expanded in spherical harmonic functions up to degree $\ell_{max}$ in colatitude $\theta$ and longitude $\phi$ and in Chebyshev polynomials up to degree $N_r$ along the radius. The time integration is performed using the ARS343 semi-implicit time scheme \citep{ascher97,gopinath22}.
We build a dataset of $21$ numerical simulations, with a fixed Prandtl number of one and covering the parameter range $10^{-6} \leq E \leq 10^{-4}$ and $ 10^7 \leq Ra \leq 2 \times 10^{10}$. The full set of simulations and related numerical truncation are given in Table~\ref{tab:NumSimus}. To mitigate the numerical cost and allow an exploration of lower Ekman numbers, the radius ratio is set to $\eta = 0.8$, which lies in the low range expected for the subsurface oceans in the Solar System (see Table~\ref{tab:Annexe1}).
\begin{table*}[h!]\centering
   \caption{Summary of the 21 simulations performed in this study.  
   $E$ and  $Ra$ are the input Ekman and Rayleigh numbers, respectively. 
   $N_r \times \ell_{max}$ defines the numerical truncation. 
   The five rightmost columns feature diagnostic parameters such as the Rossby number $Ro$, the energy ratios $E_Z/E_T$, $E_Z^g/E_Z$, $E_{R}^g/E_R$, and the zonostrophy index $R_\beta$ (see text for details).  \label{tab:NumSimus}}
 \begin{tabular}{r r r r r r r r r}
 \toprule
N$^{\circ}$ & $E$ & $Ra$ & $N_r \times \ell_{max}$ & $Ro$ & $\frac{E_Z}{E_T}$ & $\frac{E_Z^g}{E_Z}$ & $\frac{E_{R}^g}{E_{R}}$ & $R_{\beta}$ \\ [0.5ex]
\midrule
1 & $10^{-6}$ & $1 \times 10^{9}$ & 257 $ \times $ 682 & $6.3 \times 10^{-4}$ & 0.17 & 0.99 & 0.85 & 1.18 \\ [0.5ex]
2 & $10^{-6}$ & $2 \times 10^{9}$ & 321 $ \times $ 1024 & $1.6 \times 10^{-3}$ & 0.38 & 0.99 & 0.77 & 1.37 \\ [0.5ex]
3 & $10^{-6}$ & $2.5 \times 10^{9}$ & 321 $ \times $ 1024 & $2.0 \times 10^{-3}$ & 0.43 & 0.99 & 0.77 & 1.39 \\ [0.5ex]
4 & $10^{-6}$ & $3 \times 10^{9}$ & 385 $ \times $ 1365 & $2.4 \times 10^{-3}$ & 0.42 & 0.99 & 0.70 & 1.42 \\ [0.5ex]
5 & $10^{-6}$ & $5 \times 10^{9}$ & 385 $ \times $ 1365 & $4.0 \times 10^{-3}$ & 0.46 & 1.00 & 0.64 & 1.44 \\ [0.5ex]
6 & $10^{-6}$ & $1 \times 10^{10}$ & 513 $ \times $ 1365 & $6.7 \times 10^{-3}$ & 0.36 & 0.99 & 0.59 & 1.40\\ [0.5ex]
7 & $10^{-6}$ & $2 \times 10^{10}$ & 705 $ \times $ 1365 & $1.1 \times 10^{-2}$ & 0.30 & 0.99 & 0.47 & 1.34  \\ [0.5ex]
\midrule
8 & $10^{-5}$ & $5 \times 10^{7}$ & 129 $ \times $ 512 & $2.7 \times 10^{-3}$ & 0.07 & 0.97 & 0.84 & 1.09 \\ [0.5ex]
9 & $10^{-5}$ & $1 \times 10^{8}$ & 129 $ \times $ 554 & $5.8 \times 10^{-3}$ & 0.13 & 0.99 & 0.73 & 1.16 \\ [0.5ex]
10 & $10^{-5}$ & $1.3 \times 10^{8}$ & 129 $ \times $ 597 & $7.5 \times 10^{-3}$ & 0.15 & 0.99 & 0.67 & 1.15 \\ [0.5ex]
11 & $10^{-5}$ & $1.5 \times 10^{8}$ & 129 $ \times $ 597 & $8.5 \times 10^{-3}$ & 0.15 & 0.99 & 0.64 & 1.15 \\ [0.5ex]
12 & $10^{-5}$ & $1.8 \times 10^{8}$ & 129 $ \times $ 597 & $1.0 \times 10^{-2}$ & 0.17 & 0.99 & 0.61 & 1.16 \\ [0.5ex]
13 & $10^{-5}$ & $2 \times 10^{8}$ & 161 $ \times $ 597 & $1.1 \times 10^{-2}$ & 0.17 & 0.99 & 0.59 & 1.15 \\ [0.5ex]
14 & $10^{-5}$ & $5 \times 10^{8}$ & 193 $ \times $ 682 & $2.2 \times 10^{-2}$ & 0.24 & 0.99 & 0.48 &  1.18 \\ [0.5ex]
15 & $10^{-5}$ & $1 \times 10^{9}$ & 257 $ \times $ 853 & $3.6 \times 10^{-2}$ & 0.26 & 0.99 & 0.41 & 1.18 \\ [0.5ex]
16 & $10^{-5}$ & $2 \times 10^{9}$ & 321 $ \times $ 1024 & $5.4 \times 10^{-2}$ & 0.20 & 0.98 & 0.37 & 1.14  \\ [0.5ex]
\midrule
17 & $10^{-4}$ & $5 \times 10^{6}$ & 97 $ \times $ 256 & $1.9 \times 10^{-2}$ & 0.05 & 0.95 & 0.66 & 0.98 \\ [0.5ex]
18 & $10^{-4}$ & $1 \times 10^{7}$ & 97 $ \times $ 256 & $3.4 \times 10^{-2}$ & 0.05 & 0.93 & 0.51 & 0.94 \\ [0.5ex]
19 & $10^{-4}$ & $3 \times 10^{7}$ & 97 $ \times $ 256 & $7.1 \times 10^{-2}$ & 0.03 & 0.86 & 0.39 & 0.95 \\ [0.5ex]
20 & $10^{-4}$ & $1 \times 10^{8}$ & 193 $ \times $ 512 & $1.4 \times 10^{-1}$ & 0.02 & 0.75 & 0.32 & 0.93 \\ [0.5ex]
21 & $10^{-4}$ & $3 \times 10^{8}$ & 257 $ \times $ 618 & $2.6 \times 10^{-1}$ & 0.04 & 0.77 & 0.29 & 0.93\\ [0.5ex]
  \bottomrule
 \end{tabular}
\end{table*}
\subsection{Diagnostics}\label{sec:Diagnostic parameters}
In order to assess the influence of the various control parameters on the global flow properties, we define several diagnostic quantities. We adopt the following notations regarding different averaging procedures. Overbars $\overline{\cdots}$ correspond to a time average, $\langle \cdots \rangle$  to a spatial average over the whole volume and $\langle \cdots \rangle_\phi$ to an azimuthal average
\begin{equation}
    \overline{f} = \frac{1}{\tau} \int_{t_o}^{t_o+\tau} f \mathrm{d} t, \ \langle f \rangle = \frac{1}{V} \int_V f \mathrm{d}V,\ 
    \langle f\rangle_\phi = \dfrac{1}{2\pi}\int_{0}^{2\pi} f\mathrm{d}\phi,
    \label{eq:operators}
\end{equation}
where $\tau$ is the time averaging interval and $V$ is the volume of the spherical shell.

The dimensionless total kinetic energy $E_T$ is defined by
\begin{equation}
    E_T = \frac{1}{2} \overline{\langle u^2 \rangle} = \sum_{\ell=1}^{\ell_{max}} \sum_{m=0}^{\ell} \overline{\varepsilon_{\ell}^m},
\end{equation}
where $\varepsilon_{\ell}^m$ is the dimensionless kinetic energy density at  spherical harmonic degree $\ell$ and order $m$. A typical dimensionless flow velocity is given by the time-averaged Rossby number,
defined as 
\begin{equation}
Ro = \sqrt{2 E_T}. 
\end{equation}
In the context of spherical rotating turbulence, it is relevant to distinguish the axisymmetric flow (or zonal flow hereafter) from its residual, non-axisymmetric counterpart.
We hence define the dimensionless zonal kinetic energy and its associated Rossby number
\begin{equation}
    E_Z=\dfrac{1}{2 V}\int_V \overline{\langle u_\phi \rangle_\phi^2} \mathrm{d}V = \sum_{\ell=1}^{\ell_{max}} \overline{\varepsilon_{\ell}^{0}}  \quad \text{and} \quad Ro_Z = {\sqrt{2 E_Z}},
\end{equation}
where the contribution of the non-axisymmetric modes ($m\neq 0$) are excluded. The dimensionless residual kinetic energy and its associated Rossby number are defined by
\begin{equation}
    E_R = \sum_{\ell=1}^{\ell_{max}} \sum_{m=1}^{\ell}  \overline{\varepsilon_{\ell}^{m}} \quad \text{and} \quad Ro_R = {\sqrt{2 E_R}},
\end{equation}
where the contribution of the axisymmetric mode ($m= 0$) is excluded. 
Qualitatively, the residual energy represents a wealth of waves, eddies, and convective instabilities, while the zonal energy pertains to large-scale axisymmetric features, commonly referred to as jets. 

Another key feature of rotating turbulence is the nearly axial flow invariance along the rotation axis $\mathbf{e}_z$, known as flow geostrophy. The geostrophic flow $\mathbf{u}^g$ 
is obtained by averaging the velocity 
along the $z$-direction  over the spherical fluid depth following
\begin{equation}
    \mathbf{u}^g(s,\phi) = \frac{1}{h^{+}(s)-h^{-}(s)} \int_{h^-(s)}^{h^+(s)} \mathbf{u}(s,\phi,z) \mathrm{d}z, 
    \label{eq:vgeos}
\end{equation}
where $(s,\phi,z)$ are the standard cylindrical coordinates. In the above expression,
$h^+(s)=-h^-(s)=\sqrt{r_o^2-s^2}$ if $s \geq r_i$, while $h^-(s)=\sqrt{r_i^2-s^2}$ and
$h^+(s)=\sqrt{r_o^2-s^2}$ if $s < r_i$. The separation between these two branches marks the location of the so-called tangent cylinder, the imaginary cylinder 
 that circumscribes the inner boundary of the domain and is parallel to $\mathbf{e}_z$. 
 The fluid depth $h(s)$ is defined as
 \begin{equation}
     h(s) = h^+(s) - h^-(s).
     \label{eq:depth}
 \end{equation}
 For the sake of simplicity, geostrophic velocities inside the tangent cylinder are computed in the Northern hemisphere only.
Following the same procedure with $\mathbf{u}^g$ than with $\mathbf{u}$, we compute the geostrophic component of the total, zonal and residual kinetic energy denoted as $E^g_T$, $E^g_Z$,$E^g_R$. 
Table~\ref{tab:NumSimus} features the total Rossby number $Ro$ and the energy ratios E$_Z$/E$_T$, E$_Z^g$/E$_Z$, E$_{R}^g$/E$_{R}$ for the 21 simulations performed in this study.

Last but not least, an estimate of the energetic forcing in the spherical convective shell is given by the mean buoyancy power averaged over the volume $P$,
\begin{equation}
    P = \frac{Ra E^2}{Pr} \overline{\langle g u_r T \rangle}, \label{eq:convPow}
\end{equation}
that corresponds to  the energy flux injected in the fluid layer by convective instabilities. 

\section{Results}\label{sec:Numerical results}
\subsection{Global properties and flow visualisation}
\begin{figure*}[h!]
  \centering
  \includegraphics[width=\textwidth]{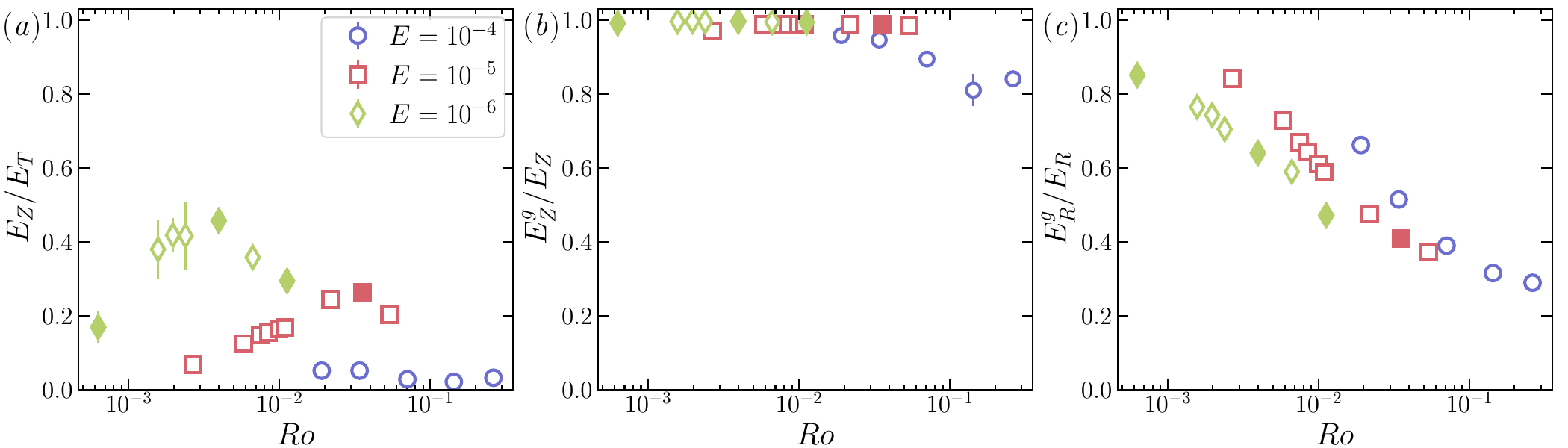}
\caption{(\textit{a}) Ratio of zonal to total kinetic energy as a function of the Rossby number $Ro$. (\textit{b}) Ratio of zonal geostrophic kinetic energy to zonal kinetic energy as a function of $Ro$. (\textit{c}) Ratio of residual geostrophic kinetic energy to residual kinetic energy as a function of $Ro$. The error bars correspond to one standard deviation about the time-averaged values. They are smaller than the symbol size for most numerical models. The filled-in symbols correspond to the simulations later highlighted in Figs.~\ref{fig:Spheres}, \ref{fig:ZonalMaps} and \ref{fig:pizzas}.
    \label{fig:IntgQuan} 
    }
\end{figure*}

Figure~\ref{fig:IntgQuan} shows energy ratios as a function of the global Rossby number 
$Ro$ across our series of $21$ numerical simulations. As previously shown by \cite{christensen02} and \cite{yadav16}, the ratio of the zonal to total kinetic energy $E_Z$/$E_T$ at a given Ekman number, follows a bell-shaped function where the maximum value increases with decreasing Ekman numbers (Fig.~\ref{fig:IntgQuan}a).
For $E = 10^{-6}$ the fraction of zonal energy reaches up to $\sim$ 50\% of the total energy, a value significantly smaller than the ratio attained in the numerical models
with stress-free boundaries \citep[see][their Fig.~3]{yadav16}.
When the Rossby number is low, it is associated with decreased Rayleigh numbers, causing the convective motions to approach a weakly non-linear regime, which is less effective in driving vigorous flows. 
On the contrary, at the largest $Ro$, a regime of intense convection is achieved, and residual energy dominates over its zonal counterpart. 
The bell-shaped pattern of the energy ratio $E_Z/E_T$ is also evident in Fig.~\ref{fig:IntgQuan}a for simulations with $E = 10^{-5}$. However, in these simulations, the maximum fraction of zonal energy does not exceed $\sim 26 \%$ of 
the total energy.
Ultimately, simulations conducted with the largest Ekman number ($E = 10^{-4}$) are strongly dominated by turbulent fluctuations, with $E_Z/E_T$ that remains below $10$\%. 
 For this Ekman number, viscous effects play a too significant role 
and preclude the formation of strong zonal jets.

We also show in Figs.~\ref{fig:IntgQuan}(b-c) the degree of  geostrophy of the zonal and residual flows. The ratio $E_Z^g/E_Z$ indicates that the zonal flow is 
almost purely geostrophic for $Ro < 10^{-2}$ (Fig.~\ref{fig:IntgQuan}b). 
On the contrary, the ratio  $E_R^g/E_R$ continuously decreases with increasing $Ro$. 
Consequently, the residual flow is partly geostrophic, and when $Ro > 10^{-2}$, half of its kinetic energy is non-geostrophic.
\begin{figure*}[h!]
  \centering
  \includegraphics[width=0.8\textwidth]
    {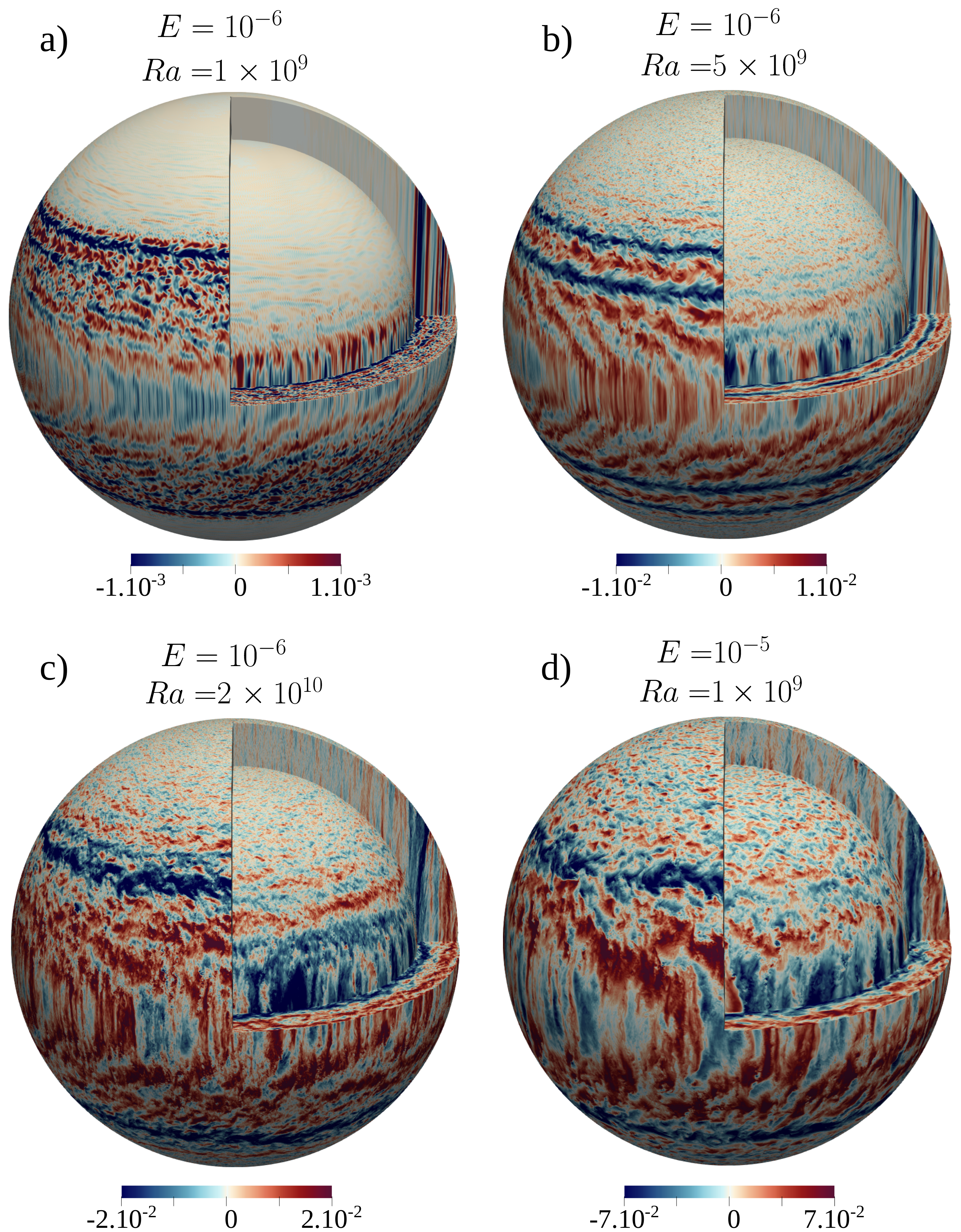}
    \caption{Instantaneous zonal velocity maps in units of Rossby number, at  $r=4.98$ for the outermost spherical surface and $r=4.05$ for the innermost one. 
    The horizontal cut corresponds to the equatorial plane. Panels a-d correspond to simulations~1, 5, 7 and 15 in Table~\ref{tab:NumSimus}.
    \label{fig:Spheres} 
    }
\end{figure*}

\begin{figure*}[h!]
  \centering
  \includegraphics[width=0.8\textwidth]
    {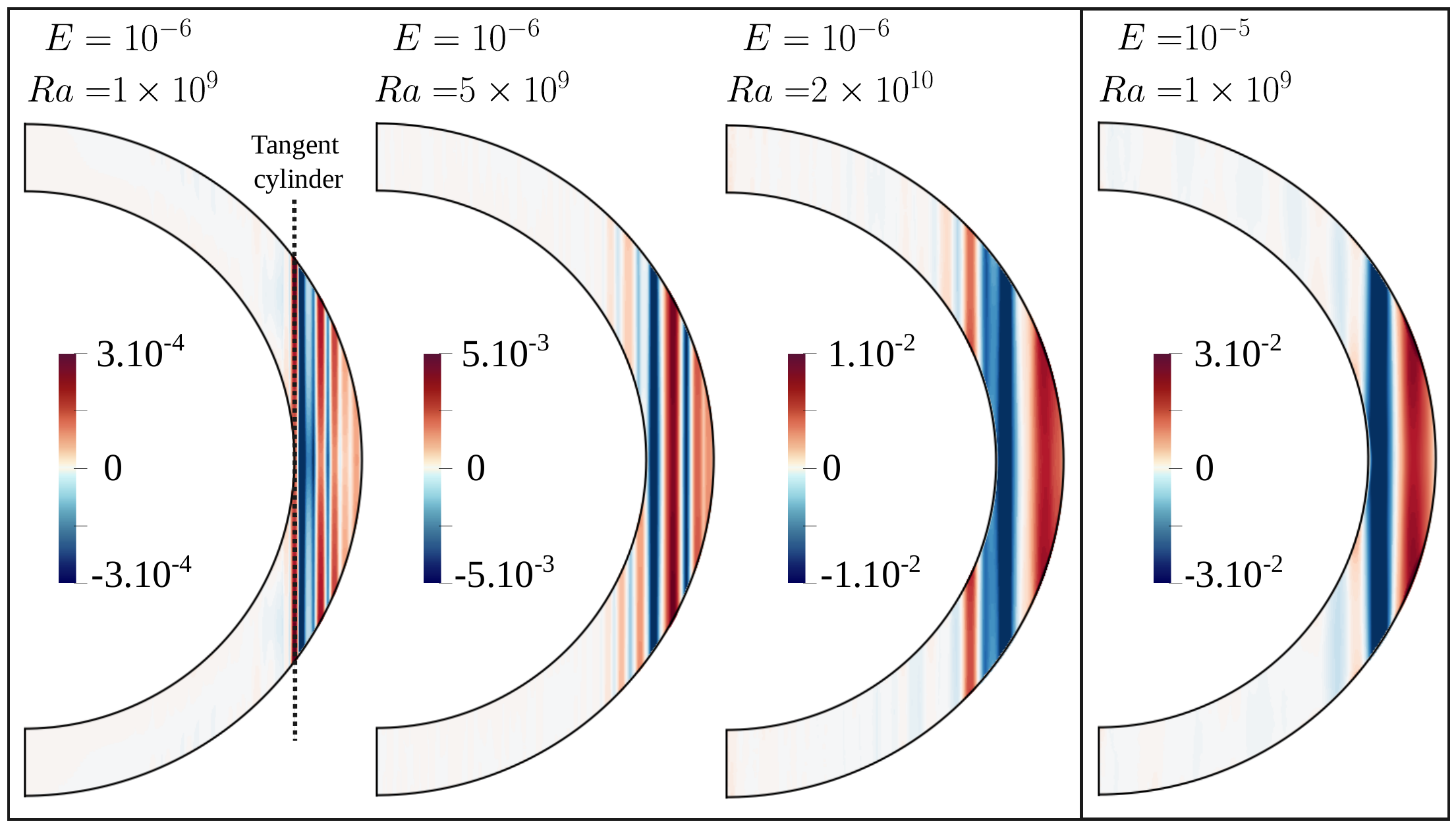}
    \caption{Time and azimuthal average of the azimuthal velocity in units of Rossby number, for the four simulations shown in  Fig.~\ref{fig:Spheres}.
    \label{fig:ZonalMaps} 
    }
\end{figure*}

To illustrate the diversity of our numerical simulations, we now show 
in Fig.~\ref{fig:Spheres} equatorial, meridional and radial cuts of the dimensionless azimuthal velocity $u_\phi$, that correspond to snapshots extracted from three 
selected cases at $E = 10^{-6}$ and one case at $E = 10^{-5}$. 
The filled-in symbols in Fig.~\ref{fig:IntgQuan} mark their location in terms of energy ratios. 
In all four images the flow is split into two dynamical regions, separated by the tangent cylinder:
a polar region (inside the tangent cylinder) which is dominated by the formation of small-scale convective plumes, and an equatorial region (outside the tangent cylinder) 
which features azimuthally-elongated structures that correspond to prograde (in red) and retrograde (in blue) zonal jets.
To better characterize this multiple jets system, we also report in Fig.~\ref{fig:ZonalMaps} the time and azimuthal averages of $u_\phi$ in the meridional plane. 
If the jets are essentially confined outside the tangent cylinder (i.e. in the equatorial region), Fig.~\ref{fig:ZonalMaps} reveals that increasing the Rossby number tends to enlarge their size and enable the formation of low-latitude jets inside the tangent cylinder. 
Interestingly, Figure~\ref{fig:Spheres}b also reveals the presence of large-scale chevron patterns in the equatorial region.
Such patterns are reminiscent of planetary Rossby waves detected in the Sun \citep{gizon21} and further characterized in a numerical model
of solar-like rotating convection \citep{bekki22}.
As shown in Fig.~\ref{fig:Spheres}b, Simulation 5, in which zonal and residual energy are almost in equipartition, provides 
a perfect illustration of the coexistence of a prograde equatorial jet and an embedded large-scale wave.
These features are less clear in the other simulations where zonal flows never account for more than 30\% of the total energy.

\begin{figure*}[h!]
  \centering
  \includegraphics[width=\textwidth]
    {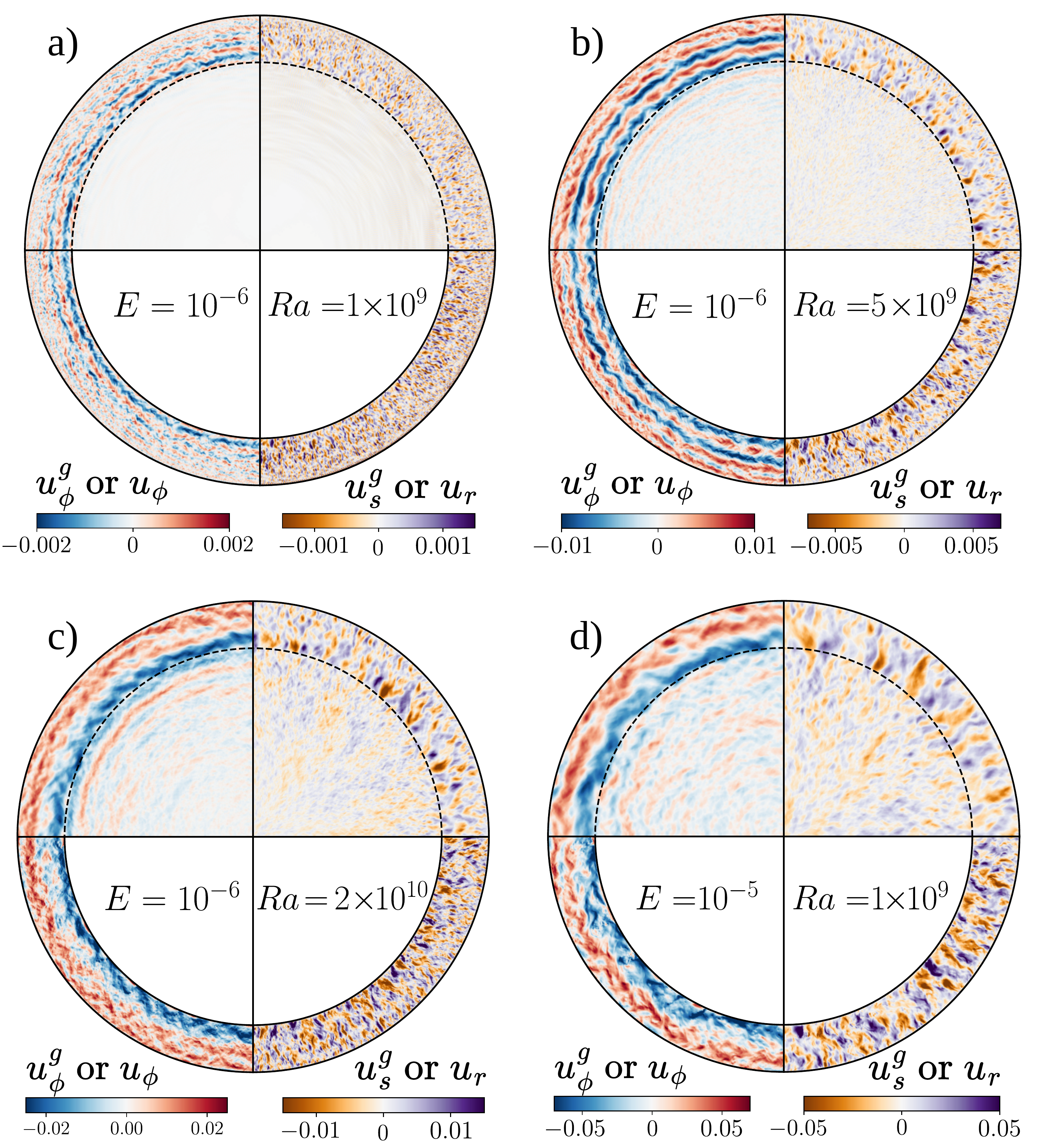}
\caption{Snapshots of different flow components for the four simulations shown in
Fig.~\ref{fig:Spheres} and \ref{fig:ZonalMaps}. For each panel, the upper half corresponds to the geostrophic flow components with $\up^g$ on the left and $u_s^g$ on the right, while the bottom half corresponds to the flow in the equatorial plane with $\up(\theta=\pi/2)$ on the left and $u_r(\theta=\pi/2)$ on the right. The dashed arcs in the upper halves mark the location of the tangent cylinder.}
    \label{fig:pizzas} 
\end{figure*}

To further illustrate the degree of geostrophy of the different flow 
components and the regionalization of the dynamics, Fig.~\ref{fig:pizzas} 
shows a comparison between $u_r$ and $\up$ in the equatorial plane (lower 
halves) with their geostrophic counterparts (upper halves). Increasing
the convective forcing at a given Ekman number (panels a to c) goes along
with broader jets and larger-scale convective features outside the tangent cylinder. While there is hardly any convective flow inside the tangent cylinder at $Ra=10^9$ indicating sub-critical polar convection \citep{gastine23}, the differences in flow amplitude gradually taper off at large supercriticalities ($Ra\approx 150\,Ra_C$ in both cases shown in panels c and d).  
The comparison of equatorial and geostrophic flows reveals that the geostrophic averaging mostly implies smoothing out the small-scale eddies while preserving the large scale structures 
\citep[for a similar analysis in geodynamo models, see Fig.~5 in][]{schwaiger21}. 
As already shown in Fig.~\ref{fig:IntgQuan}c, the global ageostrophic residual flow contributions grow with the Rossby number, which translates into increased differences between equatorial and geostrophic flows from panel a to panel d.

As a preliminary conclusion, a general picture of our simulations is that thermal convection, in a rotating spherical shell subjected to rigid mechanical boundaries, tends to spontaneously develop powerful multiple geostrophic zonal flows of alternated directions \citep[for other examples see also][]{guervilly17,barrois22,gastine23}. This trend becomes more pronounced as the Rossby and Ekman numbers are reduced. 
However, within the range of parameters achieved in the present study, the residual energy remains the predominant component of the flow, at the expense of the zonal energy.
Reaching a configuration where the zonal energy prevails over the residuals
would necessitate Ekman numbers lower by at least one order of magnitude, presently out of reach to 3D computations if one wishes to conduct a systematic parameter survey.

\subsection{Statistical analysis}
As stated in the introducion, the theoretical framework of zonostrophic turbulence was developed to study 2D rotating flows in the presence of a $\beta$-effect. It has recently been applied to mechanically driven laboratory flows, shallow-water numerical experiments and atmospheric observations. Here, we aim to extend its applicability to 3D convection in a rotating spherical shell. Consequently, a spectral analysis of the flow is in order.

\subsubsection{Theoretical energy spectra}
In order to account for the zonal anisotropy of the flow, the framework of zonostrophic turbulence  defines a 
zonal spectrum and a residual spectrum in the following universal form \citep{sukoriansky02,galperin10},
\begin{subequations}
\label{eq:optim}
\begin{align}
   E_R(k) &= C_R \Pi^{2/3} k^{-5/3} \label{eq:ER}\\
   E_Z(k_s) &= C_Z \beta^2 k^{-5}.   \label{eq:EZ}
\end{align}
\end{subequations}
The residual spectrum \eqref{eq:ER} closely follows the classical Kolmogorov-Batchelor-Kraichnan (KBK) theory of 2D isotropic turbulence, where $k$ is the total wavenumber, $C_R \sim 5-6$ is taken to be a universal constant and $\Pi$ is the energy transfer rate between the different scales of motions \citep{boffetta12}. 
The zonal spectrum \eqref{eq:EZ} characterizes the axisymmetric energy and is measured along wavenumbers $k$ in the direction orthogonal to the zonal flow. 
$C_Z$ is assumed to be a universal constant of order unity, whose value was shown to lie around $0.5$ in 
numerical simulations on the sphere \citep{sukoriansky02, sukoriansky07}, around $2$ for Jupiter \citep{galperin14}, and in the range $0.3-2.7$ for the laboratory experiments described in \cite{cabanes17} and \cite{lemasquerier23}. 
The $\beta$ parameter represents the latitudinal gradient of planetary vorticity. 
In 2D spherical flows, it directly stems from the variation of the Coriolis parameter with the colatitude $\theta$ and can be 
expressed as $\beta = 2\Omega |\sin \theta| /r_o$. In 3D spherical shells, 
$\beta$  arises from the variations of the fluid layer depth with the cylindrical 
radius $s$ and is commonly referred to as the topographic-$\beta$ \citep[e.g.][]{heimpel07}.
It is then expressed by 
\begin{equation}
\beta(s) = \frac{2\Omega}{h}  \frac{\mathrm{d}h}{\mathrm{d}s},
\end{equation}
with $h(s)$ the fluid depth defined in Eq.~\eqref{eq:depth}. To compare theoretical predictions with the velocity fields obtained from numerical simulations, it is necessary to conduct a spectral decomposition of the flow.

\subsubsection{Spectral flow decomposition}
Given the spherical geometry of our numerical setup, 
spherical harmonic functions could spontaneously 
 be invoked to compute energy spectra along the 
$\theta$ and $\phi$ directions \cite[see for example][]{boer83,boning23}. However, rotationally-constrained flows show a pronounced invariance along
the axis of rotation, as shown in Figs.~\ref{fig:Spheres} and \ref{fig:ZonalMaps}. 
In such flows, dominant features are cylindrical in nature, which favors Bessel-Fourier basis functions for spectral analysis in the $s$ and $\phi$ directions, respectively. This choice impacts the typical scales that emerge from the spectral decomposition. As an illustrative example, when a velocity field is projected onto the spherical surface, the equatorial jet appears significantly larger in comparison to the jets at higher latitudes (see Fig.~\ref{fig:Spheres}). However, when observed along the cylindrical radial direction (as shown in the meridional cross-sections in Fig.~\ref{fig:ZonalMaps}), all jets appear to have the same width, which allows in turn to unambiguously relate a wavenumber with a lengthscale. 

To carry out a spectral flow decomposition in cylindrical geometry, we make use of two distinct velocity fields: the geostrophic velocity projected onto a disc (see Eq.~\ref{eq:vgeos}), and the velocity on the equatorial annulus at $\theta=\pi/2$ (see the equatorial cut in Fig.~\ref{fig:Spheres}).
As detailed in Section~\ref{sec:Diagnostic parameters}, the geostrophic velocity is defined on the disc, at all cylindrical radii, by averaging along $\mathbf{e}_z$, while the equatorial velocity is only defined on the annulus outside the tangent cylinder.
In both cases, the domain is periodic in $\phi$ and is finite in radius, with $s\in[a,b]$, say. 
Following \cite{wordsworth08} and \cite{lemasquerier23}, one can then decompose any 
field $f(s,\phi)$ defined over such a domain into spectral coefficients using Bessel-Fourier transforms,
\begin{equation}
    \hat{\hat{f}}_{mn} =\int_a^{b} \int_0^{2\pi} f(s,\phi) \Psi_{mn}(s) e^{-im\phi} s\,\mathrm{d}s\,\mathrm{d}\phi,
    \label{eq:bessel_fourier}
\end{equation}
with $m$ and $n$ the azimuthal and radial dimensionless 
wavenumbers, respectively \citep[for a review of different spectral decompositions on the disk, see][]{boyd11}. 

On the disc, the integral along the cylindrical radius $s$ is defined over 
the interval $[0,r_o]$, and $\Psi_{mn}(s)\equiv J_m( \alpha_{mn} s )$, where $J_m$ denotes the Bessel function of the first kind of order $m$. For a disc subjected to a Dirichlet boundary condition \citep[e.g.][]{sneddon1946}, the wavenumbers $\alpha_{mn}$ are the roots of the equation
\begin{subequations}\label{eq:Jmn0}
\begin{align}
   J_m( \alpha_{mn} r_o )=0\,.
\end{align}
\end{subequations}
The radial wavenumbers $k_{mn} = \alpha_{mn}$ take discrete values since the domain is bounded \citep[see for example][]{wang09}. This spectral decomposition defines the Hankel transform on
the disc \citep[e.g.][]{baddour19}.

On the annulus $s\in[r_i,r_o]$, the support functions consist of a linear combination of Bessel functions of the first and second kind \citep{macrobert32}
 $J_m$  and $Y_m$ such that 
\begin{equation}
\Psi_{mn}(s)\equiv Y_m(\alpha_{mn} r_o) J_m(\alpha_{mn} s) - J_m(\alpha_{mn} r_o) Y_m(\alpha_{mn} s)\,.
\end{equation}
When Dirichlet boundary conditions are enforced on both sides, the 
wavenumbers $\alpha_{mn}$ are the roots of 
\citep{cinelli65}
\begin{equation}
Y_m(\alpha_{mn} r_i) J_m(\alpha_{mn} r_o) - J_m(\alpha_{mn} r_i) Y_m(\alpha_{mn} r_o)=0\,.
\end{equation}
For this combination of geometry and boundary conditions, Eq.~\eqref{eq:bessel_fourier} is then known as the Weber-Orr transform.

We compute kinetic energy spectra following
\begin{equation}
    E_{mn} =  \dfrac{r_o}{2V^\star}  N_{mn}\left(|\mathcal{U}_{mn}|^2+|\mathcal{V}_{mn}|^2\right)\,,
    \label{eq:e_mn}
\end{equation}
where $\mathcal{U}_{mn}\equiv \widehat{\sqrt{h} \hat{u}_s}$ and $\mathcal{V}_{mn}\equiv \widehat{\sqrt{h} \hat{u}_\phi}$ to account for the height variation of the fluid depth $h$ defined in Eq.~\eqref{eq:depth}.
Note that a single hat $\widehat{\cdots}$ corresponds to a Fourier transform, while two hats correspond to a Bessel-Fourier transform.
In the above equation, $V^\star$ either corresponds to the full fluid volume  in the case of the disk ($V^\star=V$) or to the volume outside the tangent cylinder only in the case of the annulus ($V^\star=V^a=(4\pi/3)[h^+(r_i)]^3$).
In the above expression, $N_{mn}$ is a normalisation factor which depends upon the relevant Bessel-Fourier transform such that
\begin{equation}
N_{mn}= \dfrac{2\pi}{r_o^2}\dfrac{1}{J_{m+1}^2(\alpha_{mn}r_o)}
\end{equation}
for the disc \citep[see][]{sneddon1946,guizar04,baddour19}, and
\begin{equation}
N_{mn}=\dfrac{\pi^3\alpha_{mn}^2J_m^2(\alpha_{mn}r_o)}{J_m^2(\alpha_{mn}r_i)-Y_m^2(\alpha_{mn}r_o)}
\end{equation}
for the annulus \citep[][]{sneddon1946,cinelli65}. 
 
We distinguish the zonal spectrum $E_Z(k_{0n}) = E_{0n}$ that contains the kinetic energy of the azimuthal mode $m=0$ and the residual spectrum $E_R(k_{0n}) = \sum_{\substack{m=-\ell_\text{max}, m\neq0}}^{\ell_\text{max}}  E_{mn}$, which is the contribution of all non-zonal modes, $m \neq 0$.
In this spectral analysis, each mode $m$ has different wavenumbers $k_{mn} = \alpha_{mn}$
and the associated zonal wavenumbers are $k_{0n} =  \alpha_{0n}$. As suggested by \cite{lemasquerier23} we perform a summation into spectral bins for modes $m \neq 0$ in order to compute the residual spectrum. The spectral bins $\mathrm{d}k$ are defined to correspond to the zonal wavevector $k_{0n}$. 
Here, we consider that any typical length scale is half a period in radius and can be computed using $L = \pi/k_{mn}$. Hereafter, all spectra are computed once the steady state is achieved in our simulations and a time average over at least $20$ statistically independent spectra is computed.

\subsubsection{Residual spectra}
\begin{figure*}[h!]
  \centering
  \includegraphics[width=0.7\textwidth]
    {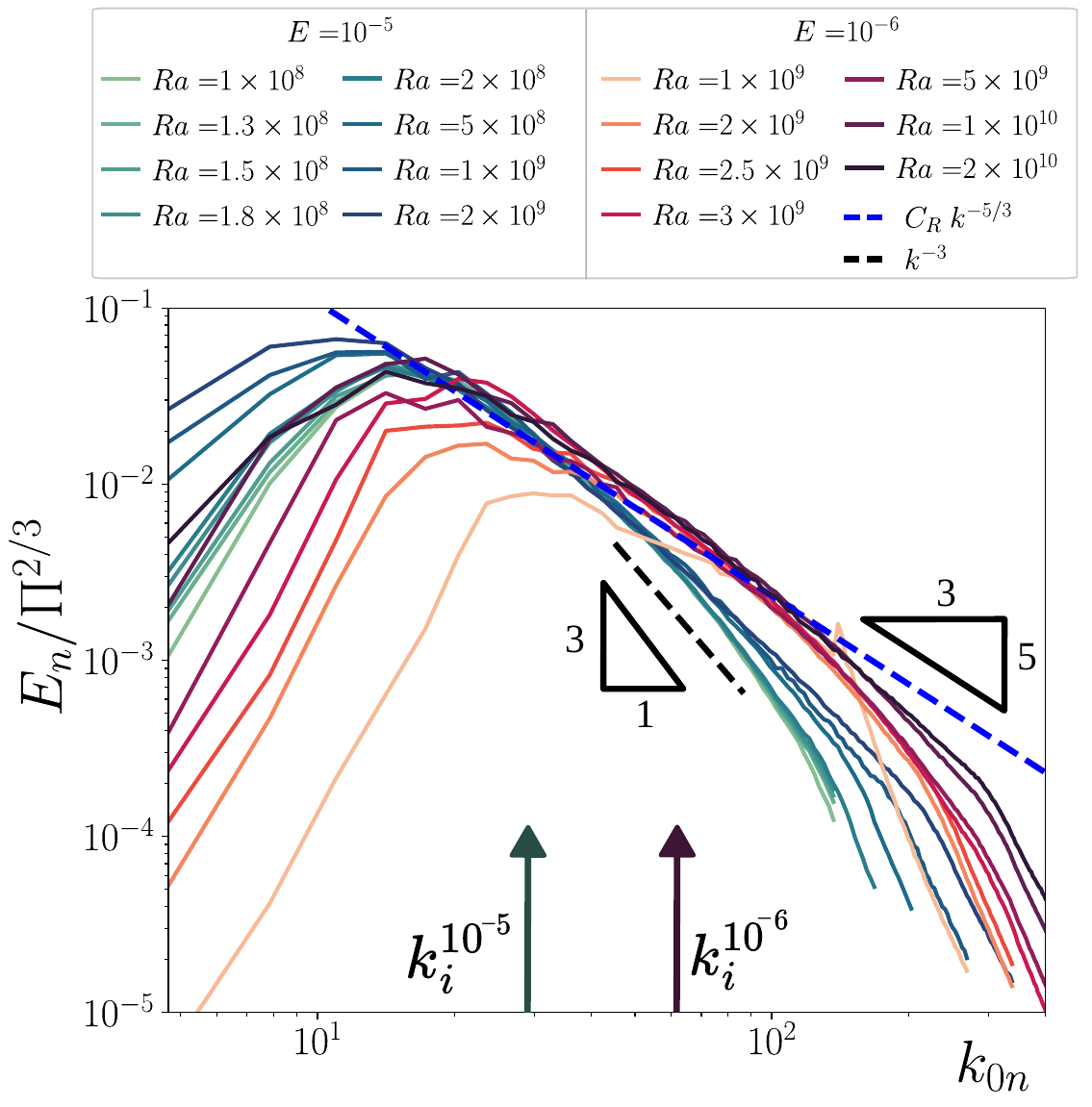}
    \caption{Residual spectra of the equatorial velocity on the equatorial annulus. The vertical arrows correspond to the two injection wavenumbers $k_i^E$ for $E=10^{-5}$ and $E=10^{-6}$. Green to blue color gradient goes from $Ra = 10^{8}$ to $ 2\times 10^{9}$ for simulations at $E = 10^{-5}$. Light orange to purple color gradient goes from $Ra = 10^{9}$ to $ 2\times 10^{10}$ for simulations at $E = 10^{-6}$. The theoretical prediction of the residual spectra from Eq.~\eqref{eq:ER} corresponds to the dashed blue line using $C_R = 5$. The dashed black line corresponds to a $k_{0n}^{-3}$ slope. 
    \label{fig:EKResiduals} 
    }
\end{figure*}

Figure~\ref{fig:EKResiduals} shows the residual energy spectra computed from the equatorial velocity on the annulus and for the entire set of simulations at $E = 10^{-5}$ and  $10^{-6}$.
At a given Ekman number, the injection scale is estimated by $k_i^E=m_c/r_i$ using the critical azimuthal wavenumber at onset of 
convection $m_c$ computed by \citet{barik23}, which give $m_c=116$ and $247$ for $E = 10^{-5}$ and $10^{-6}$, respectively.
Once the statistically-steady state has been reached, $\Pi=P$ (Eq.~\ref{eq:convPow}),
providing an effective way to measure the energy transfer rate in convection driven turbulence. This enables a straightforward comparison of the residual spectra for different convective forcings by considering $E_R(k)/P^{2/3}$.
All residual spectra are then found to follow the $-5/3$ slope expected from the KBK theory of 2D-turbulence, with $C_R \approx 5$ in Eq.~\eqref{eq:ER}.

This also shows that when the Rayleigh number increases and the Ekman number decreases, the range of scales where the $-5/3$ scaling applies increases, with larger wavenumbers exhibiting a slope near $-3$.
This change in the spectral slope occurs at wavenumbers close to the injection scale $k_i^E$.
Such flow properties are reminiscent of the paradigm of 2D-turbulence coined by \cite{Kraichnan67}: an inverse cascade of kinetic energy from injection to large scales with a $-5/3$ slope, and a direct cascade of enstrophy to small scales with a $-3$ slope (see \cite{deusebio14} and \cite{boffetta12} for a review). 
Figure~\ref{fig:EKResiduals} illustrates that
lowering the Ekman number goes along with a broadening of the range of scales that adhere to the $-5/3$ slope towards small wavenumbers, possibly suggesting a more effective inverse cascade.
Recently, \cite{boning23} showed that a $-5/3$ slope can also be found in spherical convective flows, where zonal jets are driven by statistical correlations of small convective scales instead of the inverse turbulent cascade invoked in the theory of 2D-turbulence.
To differentiate between these two mechanisms and determine which one is at play in our numerical simulations, a more detailed analysis of energy and enstrophy transfers would be necessary, an analysis that goes beyond the scope of the current study.

\begin{figure}[h!]
  \centering
  \includegraphics[width=0.49\textwidth]
    {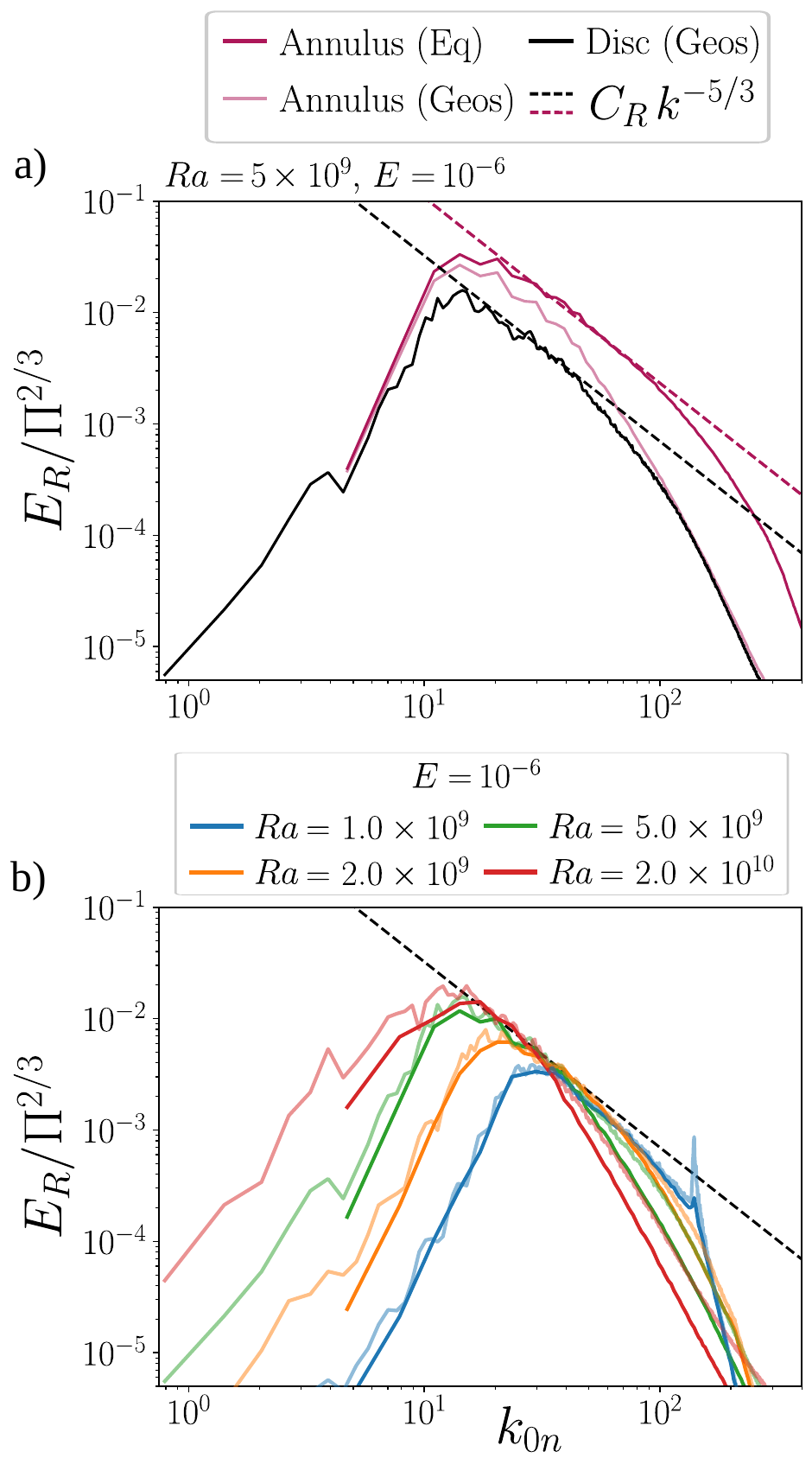}
\caption{(\textit{a}) Residual spectra of simulation 5, computed for the geostrophic (geos) and equatorial (Eq) velocities on the annulus as well as for the geostrophic velocity on the disc. The pink curve is the same as the one presented in Fig~\ref{fig:EKResiduals}. (\textit{b}) Comparison of the residual spectra computed on the disc (light lines) and on the annulus (opaque lines) for the geostrophic residal velocity for four selected simulations at $E = 10^{-6}$ with increasing supercriticalities. To stress the influence of the geometric factor, the residual spectra on the annulus have been multiplied by $V^a/V$.
On both panels, the dashed pink and black lines are theoretical predictions from Eq.~\eqref{eq:ER} with $C_R = 5$ and $1.5$ respectively. 
}
\label{fig:EKResidualsGeosEq}
\end{figure}

In Fig.~\ref{fig:EKResidualsGeosEq}a, we expand our statistical analysis by displaying three distinct residual spectra computed from simulation 5, which corresponds to the maximum energy ratio $E_Z/E_T = 0.5$ and the control parameters $E=10^{-6}$ and $Ra = 5 \times 10^9$.
In this analysis, we concurrently display the residual spectra computed on the annulus, originating from both the equatorial (pink) and geostrophic (light pink) velocities, along with the spectrum computed on the disc from the geostrophic velocity (black).
The light pink and black curves illustrate a decrease in the magnitude of the residual spectra for the geostrophic velocity when compared to the equatorial velocity. This reduction becomes even more pronounced when we consider a spectral decomposition on the disc rather than on the annulus.
It is evident that part of this reduction results from the axial averaging involved in the computation of the geostrophic component. Referring to Fig.~\ref{fig:IntgQuan}, we can recall that only 64\% of the residual energy is attributed to the geostrophic component for simulation 5 (see also Tab.~\ref{tab:NumSimus}).
However, the magnitude of the energy spectrum on the disc is further reduced compared to that on the annulus. This reduction can be attributed to the presence of the tangent cylinder, which forms a dynamical barrier leading to different regionalized dynamics between high and low latitude regions \citep[e.g.][]{gastine23}.
%

These different regions manifest as a consequence of the inherent anisotropy of the convective forcing. \cite{dormy04} 
demonstrated that the onset of convective instabilities in a rotating spherical shell is localized adjacent to, but outside, the tangent cylinder. 
Additionally, imposing no-slip boundary conditions in our numerical setup further reinforces the existence of these two dynamical regions, suppressing the formation of large-scale flows within the polar region. The combination of these effects results in significant heterogeneities in turbulent mixing throughout the fluid shell.
Consequently, when incorporating the weakly convective polar region into the spectral decomposition on the disc, there is a marked reduction in the residual spectrum. This decrease can hence be attributed to  the
combined effects of partial flow geostrophy and differences of flow amplitude inside and outside the tangent cylinder. The latter can be approximated by a
geometric factor $V/V^a\approx 2.2$ in the limit of sub-critical convection inside the tangent cylinder.
To account for those effects, we fit the black residual spectrum in Fig.~\ref{fig:EKResidualsGeosEq}a using Eq.~\eqref{eq:ER} by lowering the residual constant from $C_R=5$ to $1.5$ (as indicated by the black dashed line).
This shows that within spherical shell convection, the residual spectra cannot be uniquely described by the theory of 2D isotropic, homogeneous, turbulence that arises in unconstrained turbulent environments \citep{Kraichnan67}.
Instead, the polar and equatorial dynamical regions, when considered separately, are nearly homogeneous and must be treated accordingly. This approach is illustrated in our consideration of the equatorial region alone in Fig.~\ref{fig:EKResiduals}.

Considering the parameter range for icy satellites, where $10^{-12} \lesssim E \lesssim 10^{-10}$ and $10^{16} \lesssim Ra \lesssim 10^{24}$, 
it is plausible that the formation of zonal jets extends to higher latitudes. Such a dynamical configuration could
lead to a more homogeneous distribution of turbulence throughout the fluid volume, causing the spectrum computed on the disc to converge towards that computed on the annulus.
To test this hypothesis, we present in Fig.~\ref{fig:EKResidualsGeosEq}b residual spectra obtained for a range of simulations with increasing convective forcings from $Ra = 10^9$ (Simulation 1) to $Ra = 2\times 10^{10}$ (Simulation 7). To highlight the flow regionalization, the residual spectra on the annulus have been multiplied by the aforementioned geometric factor $V^a/V$.
While a nearly perfect overlap is observed for both spectra (i.e. on the disc and the annulus) in the weakly-forced case (blue lines), they gradually depart from each other as supercriticality increases.
This deviation is more sizeable at larger scales ($k_{0n}\leq 10$), particularly for the highly turbulent case $Ra=2\times 10^{10}$ (red lines).
This reflects a growing fraction of energy inside the tangent cylinder, as also illustrated in Fig.~\ref{fig:pizzas}. 
Nonetheless, the parameters covered in this study do not allow to reach comparable energy levels inside and outside the tangent cylinder. Would this homogenization process becomes complete within the parameter range relevant to icy satellites, one can speculate that a spectral analysis on both the disc and the annulus would yield the same result, within the geometric factor $V^a/V$ involved in Fig.~\ref{fig:EKResidualsGeosEq}b.

\subsubsection{Zonal spectra}
\begin{figure}[h!]
  \centering
  \includegraphics[width=0.49\textwidth]
    {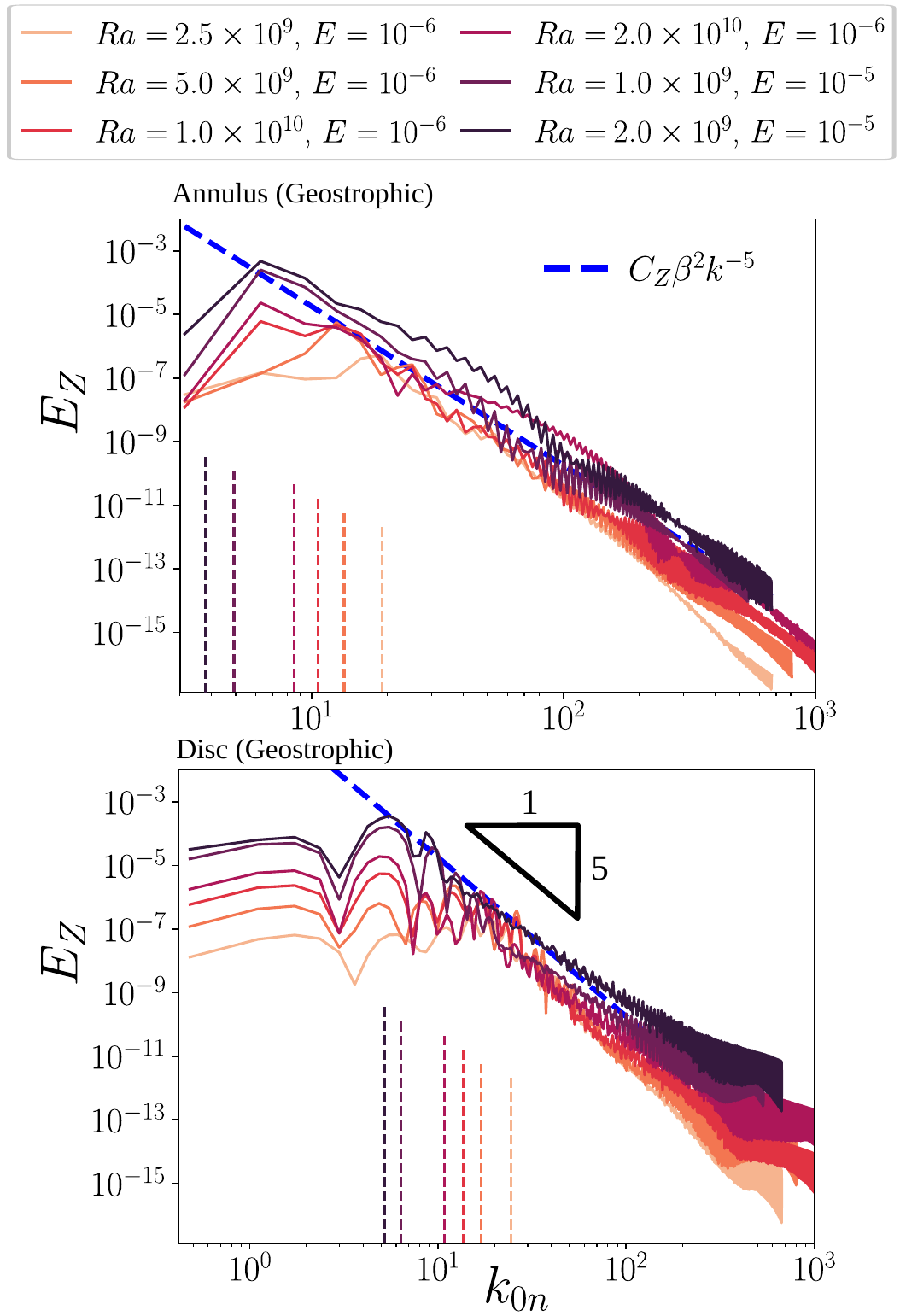}
    \caption{
     Zonal spectra of the geostrophic velocity computed on the annulus (top panel) and on the disc (bottom panel). The theoretical prediction of the zonal spectra from Eq.~\eqref{eq:EZ} corresponds to the dashed blue line using $C_Z = 0.5$. Vertical dashed segments are Rhines wavenumbers $k_{Rh}$ from Eq.~\eqref{eq:krhines} and for each simulation. \label{fig:EKzonal} 
     }
\end{figure}

Figure~\ref{fig:EKzonal} shows zonal energy spectra computed from the geostrophic velocity on the annulus and on the disc. We have selected energy spectra from different simulations at $E=10^{-6}$ and $10^{-5}$, including different Rayleigh numbers.
For all the numerical simulations, the zonal spectra consistently exhibit a $-5$ slope on approximately two decades and the energetic amplitude is well predicted by Eq.~\eqref{eq:EZ}. 
The dimensionless $\beta$ parameter is estimated at the mean radius of the equatorial annulus $s_\text{mid}$,
\begin{equation}
    \beta =  \frac{2 s_\text{mid}}{r_o^2 - s_\text{mid}^2}  \quad \text{with} \quad s_\text{mid} = \dfrac{1}{2}(r_i+r_o), \label{eq:beta}
\end{equation}
and the zonal constant which best fits the spectra is $C_Z=0.5$.
We opt for estimating $\beta$ at the mean equatorial radius because it corresponds to the average position of the zonal jets since they are mostly confined outside the tangent cylinder  (see Fig.~\ref{fig:ZonalMaps}). 
This condition generally holds for all simulations, where only weak zonal jets form in the polar region.
However, our estimate of $\beta$ may require reconsideration if zonal jets were to extend towards the poles. This is the case for instance in the simulations with free-slip boundary conditions conducted by \cite{soderlund14} or possibly in the range of parameters expected for the icy satellites.  
In such configurations, the $\beta$ parameter would need to be estimated differently to account for high latitudes.
On the other hand, we estimate the zonal constant $C_Z =0.5$ by fitting the magnitude of the zonal spectra computed on the annulus rather than on the disc.
Again, the reason for this choice is that the disc encompasses fluid regions with very heterogeneous flows. 
In addition, the spherical shell geometry has its own spectral signature which exhibits energetic bumps at intermediate wavenumbers when using Bessel-Fourier decomposition (see Fig.~\ref{fig:EKzonal}).
These bumps arise because the flow adopts the cylindrical geometry of the fluid shell (one equatorial region with strong jets, the other without), which artificially affects the energy spectra. This would probably not occur if jets would have formed at all cylindrical radii, making the spectral decomposition on the disc more relevant.
This configuration may occur in the parameter range relevant to icy satellites, or it can equally arise in simulations with free-slip boundary conditions, where significant polar and equatorial zonal flows are known to develop \citep[as observed in simulations by][]{soderlund19,boning23}. Conducting a spectral analysis of these free-slip simulations could provide further insights into the dynamics of such flows and their spectral distribution.

It is worth noting that the zonal constant $C_Z$ obtained here lies within the range of values $0.3 \lesssim C_Z \lesssim 2.7$ estimated in previous studies of zonostrophic turbulence \citep{sukoriansky02,galperin14,cabanes17,lemasquerier23}. We thus reinforce its universality by extending it to deep-seated convective flows.
Furthermore, a key implication of Eq.~\eqref{eq:EZ} is that the energetic amplitude of the zonal spectra is independent of the convective forcing and solely depends on the rotation rate $\Omega$ and some geometric parameters given in Eq.~\eqref{eq:beta}.

It is also well accepted that in flows with a strong $\beta$-effect, the Rhines scale \citep{rhines75} is a good approximation of the frictional scale denoted 
by $k_f$ in Fig.~\ref{fig:zonostrophy}.
This scale is pivotal in arresting the inverse cascade of energy in 
such flows \citep[e.g.][]{sukoriansky07}; it is defined by
\begin{equation}
    k_{Rh} = \sqrt{\frac{\beta}{2 U_g}}.
    \label{eq:krhines}
\end{equation}
where $U_g$ is the root-mean-square geostrophic velocity integrated either on the disk or on the annulus.
The dashed vertical segments in Fig.~\ref{fig:EKzonal} show that the Rhines wavenumber matches quite effectively with the most energetic scale, above which the energy systematically decreases.
The associated Rhines scale $L_{Rh} = \pi/k_{Rh}$ has also proved to be a good estimate of the typical jets width \citep[see][for examples in spherical rotating convection]{heimpel07,gastine14,bire22}.
Since the jets size increases with the Rayleigh number 
(see Figs.~\ref{fig:ZonalMaps} and \ref{fig:pizzas} for an illustration) it results in zonal spectra which further extend towards smaller wavenumbers. 
At the highest Rayleigh numbers, the energy maxima reach typical length scales 
corresponding to almost half the shell gap. As previously stated, these length scales 
cannot be clearly identified in zonal spectra computed on the disc because of the zonal 
flow regionalization.

\begin{figure}[h!]
  \centering
  \includegraphics[width=0.49\textwidth]{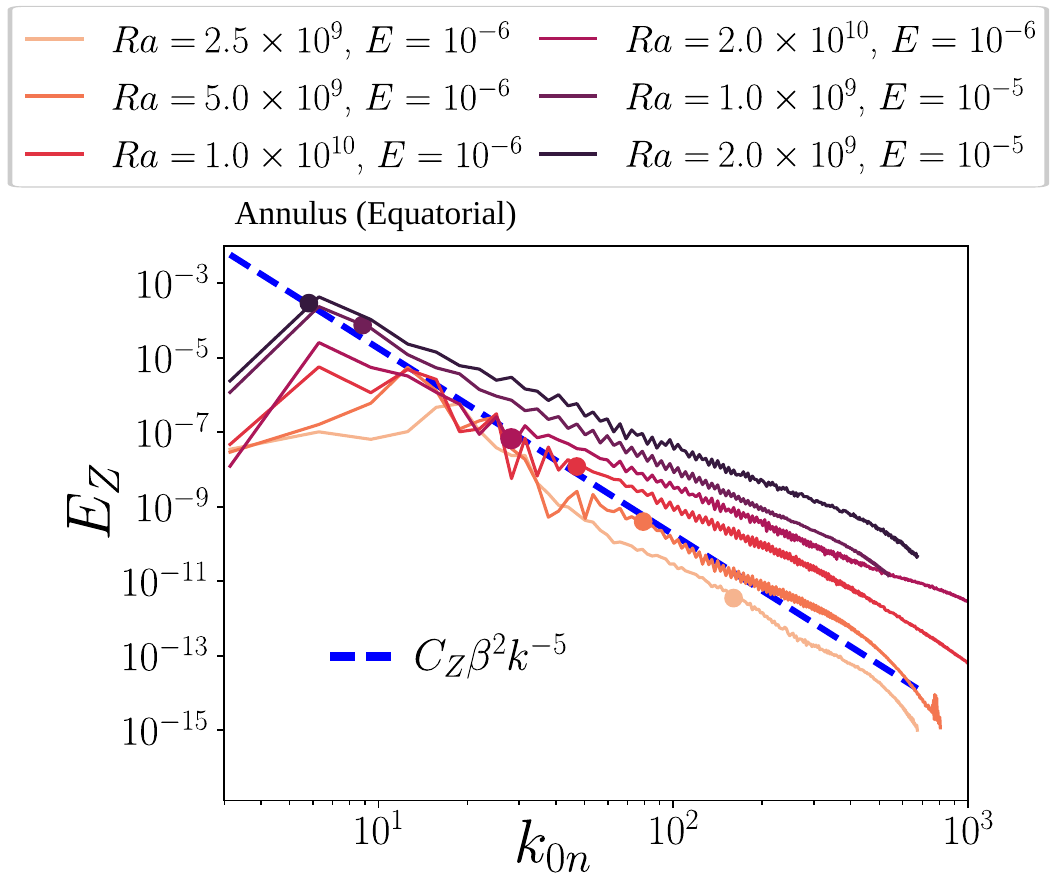}
    \caption{Zonal spectra of the equatorial velocity computed on the annulus. The theoretical prediction of the zonal spectra from Eq.~\eqref{eq:EZ} corresponds to the dashed blue line using $C_Z = 0.5$. Dots are Rossby scales $k_{Ro}$ from Eq.~\eqref{eq:kro}. 
 }
 \label{fig:EkEqZonal}
\end{figure}

Finally, we apply the zonotrophic theory to zonal energy spectra obtained from the equatorial velocity on the annulus, as shown in Fig.~\ref{fig:EkEqZonal}. Similarly to the geostrophic component, zonal energy at small wavenumbers are well accounted for by Eq.~\eqref{eq:EZ}. However, for large wavenumbers, the zonal energy of the equatorial flow shows a strong dependence on the convective forcing, since its magnitude increases with increasing Rayleigh numbers. 
The slope also departs from the $-5$ scaling and gradually tapers off on increasing convective forcings.
In order to estimate the typical wavenumber at which the zonal spectra deviate from the theoretical predictions, we introduce a Rossby wavenumber $Ro_k=Ro\,k_{0n}/\pi$.
Assuming an arbitrary threshold value of $Ro_k = 0.1$ to mark the limit of strong rotationnal constraint \citep[a similar threshold is used in the study by][]{christensen06}, we define a critical wavenumber as follows,
\begin{equation}
    k_{Ro} = \frac{0.1 \pi}{Ro}.
    \label{eq:kro}
\end{equation}
For wavenumbers where $k < k_{Ro}$, the flow is strongly geostrophic and characterized by a Rossby number lower than $0.1$. These scales are the most energetic, and as a result, global properties presented in Fig.~\ref{fig:IntgQuan} reveal a zonal flow that is predominantly geostrophic. Thus, although the theory of zonostrophic turbulence does not explain the energy distribution of the equatorial velocity at all scales, it gives a good estimate of the zonal energy in the convective flows.

Here, we would like to stress that the zonostrophic theory remains applicable in the transitional regime of rotating convection (Fig.~\ref{fig:RavsEk}),
even in setups where the zonal energy represents only a weak fraction of the total kinetic energy (see Tab.~\ref{tab:NumSimus}).
However, a crucial question arises when approaching the non-rotating limit -- a condition possibly relevant for Titan, Europa and Ganymede -- in which the flow
becomes essentially ageostrophic. 
Although our study reveals that approximately 99\% of the zonal energy is geostrophic 
for all simulations with $E=10^{-5}$ and $10^{-6}$, it is likely that this fraction decreases when trending towards the non-rotating limit.  
As the determination of the boundary of the domain where zonostrophic theory applies remains unfortunately inaccessible to present computations, it is  
difficult to provide a conclusive answer in that regard. 

For now, it is crucial to bear in mind that future energetic predictions 
for subsurface oceans, based on  zonostrophic theory, will provide an estimate for the \emph{geostrophic component} of the zonal energy.

\subsection{The zonotrophy index}
\begin{figure}[h!]
  \centering
  \includegraphics[width=0.49\textwidth]
    {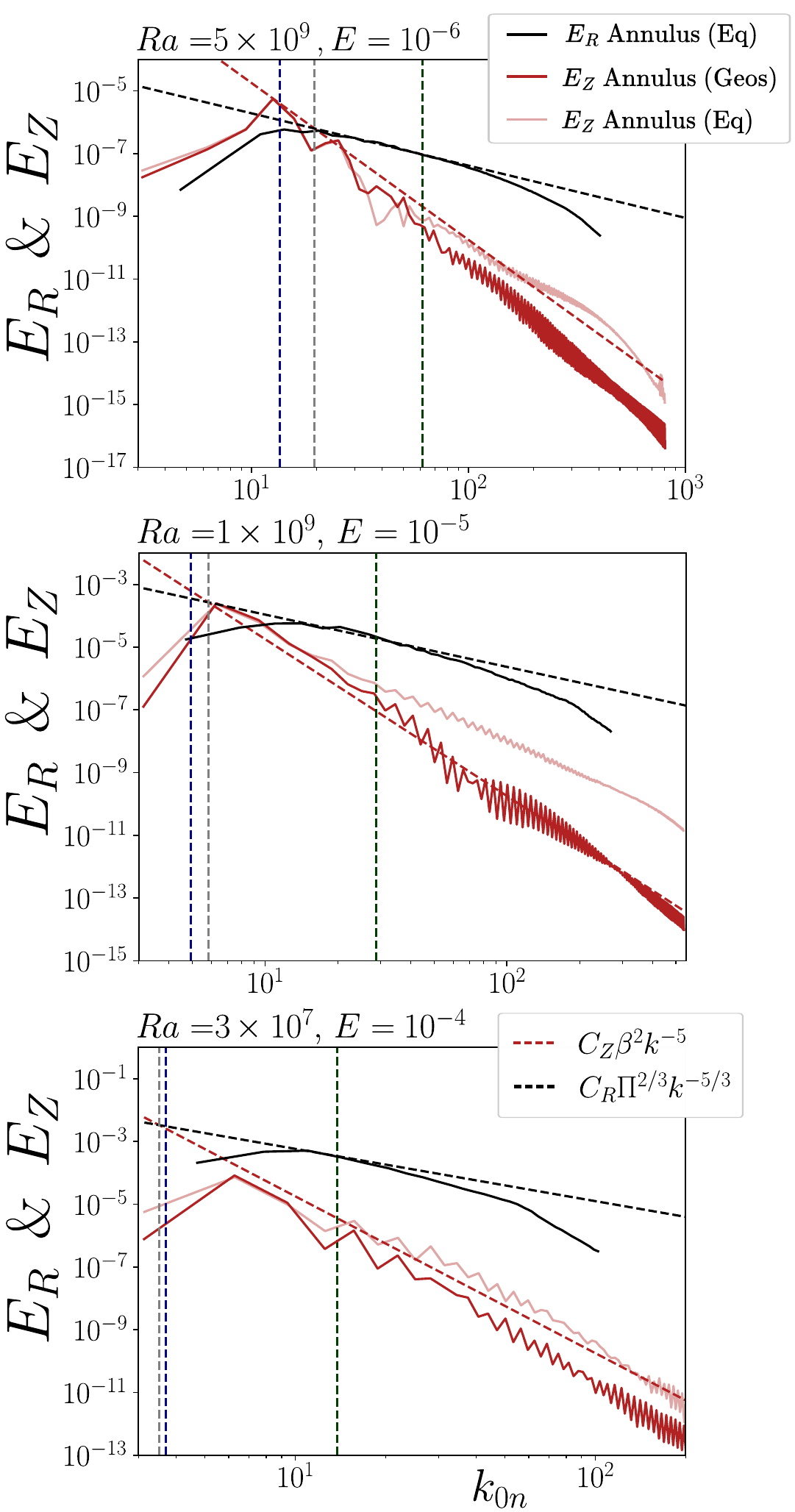}
    \caption{Zonal and residual energy spectra computed on the annulus from the equatorial (Eq) and geostrophic (Geos) velocities. Panels from top to bottom correspond to simulations~5, 15 and 19 in Table~\ref{tab:NumSimus}. The red and black dashed lines are the theoretical predictions of the zonal and residual spectra using the zonal constant  $C_Z = 0.5$ and the residual constant $C_R = 5$. Vertical lines corresponds to Rhines' wavenumber (blue), the transitional wavenumber (grey) and the injection wavenumber $k_i^E=m_c/r_i$ (green), with $m_c= 55$, 116 and 247 for $E = 10^{-4}$, $10^{-5}$ and $10^{-6}$, respectively \citep{barik23}.
    \label{fig:EREZ} 
    }
\end{figure}

We show in Fig.~\ref{fig:EREZ} the zonal and residual spectra for simulations 5, 15 and 19 corresponding to the highest ratio of zonal energy $E_Z/E_T$ for numerical experiments with $E = 10^{-6}$, $10^{-5}$ and $10^{-4}$. The intersection between the theoretical $-5/3$ (Eq.~\ref{eq:ER}) and $-5$ (Eq.~\ref{eq:EZ}) spectra defines a transitional wavenumber, which can be expressed as
\begin{equation}
    k_{\beta} = \left(\frac{C_Z}{C_R}\right)^{3/10} \left( \frac{\beta^3}{\Pi}\right)^{1/5}. \label{eq:lbeta}
\end{equation}
This wavenumber defines the scale at which the theoretical zonal spectrum overcomes the residual one. Thus, the range of wavenumbers $k_{Rh}<k<k_{\beta}$ is known as 
the zonostrophic inertial range, in which the zonal jets dominate energetically.
In rotating turbulent flows, this inertial range exists if the zonostrophy index $R_{\beta} = k_{\beta}/k_{Rh}$ is greater than unity. 
At $E = 10^{-6}$, the simulation 5 satisfies this condition with $R_{\beta} = 1.45$ and the zonal spectrum dominates the residual spectrum in the range predicted by the Rhines and transitional wavenumbers.
For the simulation 15 at $E = 10^{-5}$, the zonostrophic inertial range is significantly reduced and the intersection of the zonal and residual spectra is poorly predicted by the theory. With an index of $R_{\beta} = 1.18$, this simulation shows that the $\beta$ parameter is too weak to efficiently channel energy in the zonal flow component.  
Finally, simulation 19 at $E = 10^{-4}$ is not in the zonostrophic regime with $R_{\beta} < 1$ and the residual energy largely dominates its zonal counterpart. 
The flow is energetically governed by the convective forcing, given that the residual energy is determined by the convective power following Eq.~\eqref{eq:ER}.  
With a nearly $-5/3$ slope at all wavenumbers, the behavior of the flow resembles that of 3D-isotropic turbulence, with a forward energy cascade from energy injection at small wavenumber ($k_i^{10^{-4}}\approx 14$) down to viscous dissipation at larger wavenumber.
Nonetheless, although the residual energy dominates, the zonal spectrum preserves its $-5$ slope and magnitude predicted by Eq.~\eqref{eq:EZ}. 
In this simulation, the influence of both large viscosity ($E=10^{-4}$) and substantial rotational effects ($Ro \sim 0.07$) is evident, leading to 86\% of the zonal energy to be in geostrophic balance. Hence, the weak yet mostly geostrophic zonal flows still adhere to the $-5$ theoretical scaling.

We report in Table~\ref{tab:NumSimus} the zonostrophy indices for all numerical simulations. The general trend that emerges is that the zonostrophy index increases on decreasing Ekman numbers. This goes along with a greater fraction of zonal energy in the convective shell. Using 2D simulations on a rotating sphere, \cite{galperin10} suggest that a lower bound for the zonostrophic regime is rather $R_{\beta} \simeq 2.5$. They argue that below this value, rotating turbulence occurs in a transitional regime between the dissipation-dominated regime and zonostrophic turbulence. The threshold value of $2.5$ is however only indicative and has been established for specific cases of 2D-turbulence \citep{sukoriansky07}.
Here, in 3D convective flows, with no-slip boundary conditions at the top and bottom of the spherical shell, we never exceed the value of $R_{\beta} = 1.45$. The
laboratory experiments conducted by \cite{lemasquerier23} also reveal zonostrophy indices that do not surpass the value of $2$.
Yet, the theory accurately predicts the spectral energy distribution, even when the residual energy dominates, especially in simulations with $E = 10^{-6}$. Consistently, we argue that zonostrophic turbulence is a relevant framework for 
analysing convective flows enclosed within a rotating spherical fluid layer. We now propose to use it to make a first theoretical prediction of the typical velocity of subsurface oceans. 

\section{Implications for subsurface oceans}\label{sec:Implications for subsurface oceans}

In order to apply the framework of zonostrophic turbulence to the subsurface oceans of the icy satellites, we assume that this theory can be extrapolated to more extreme parameters, despite the distance between the planetary regimes and those covered by the 3D numerical models (recall Fig.~\ref{fig:RavsEk}). 
This extrapolation has already proven effective, as \cite{galperin14} successfully applied this theory to \textit{in situ} measurements of the jovian zonal jets collected from the Cassini and Voyager missions.
Our goal here is to use Eqs.~(\ref{eq:ER}-\ref{eq:EZ}) to generate theoretical residual and zonal spectra and estimate the corresponding velocities for the subsurface oceans of Enceladus, Titan, Europa and Ganymede.

To this end, we present in Table~\ref{tab:Annexe1} some properties of the subsurface oceans in dimensional units \cite[extracted from][]{soderlund19}: the thermal expansivity $\alpha$, the gravitational acceleration $g$, the heat flux $q$, the density $\rho$, the heat capacity $C_p$, the ocean depth $D$ and the ice thickness $D_I$.
These properties have been obtained in previous studies (see caption in Table~\ref{tab:predictions}) by considering different water compositions of the ocean, including scenarios with salty water containing 10 wt\%  MgSO$_4$, seawater, or
pure water.
Using these properties one can estimate the buoyancy power per unit mass available to drive convection in the subsurface oceans, following,
\begin{equation}
    P_s = \frac{\alpha g q}{\rho C_p}. \label{eq:buoyancy}
\end{equation}
Our estimates are given in Table~\ref{tab:predictions}.
To generate synthetic residual and zonal spectra of the subsurface oceans, we follow the same approach as in our numerical experiments:
the turbulent mixing is given by the buoyancy power $\Pi=P_s$, the $\beta$ parameter is estimated at the mean equatorial radius following Eq.~\eqref{eq:beta}, with $r_o = R-D_I$ and $r_i = r_o-D$, the residual constant is $C_R=5$ and the lower and upper bounds of the zonal constant are considered to be $0.3  \lesssim  C_Z  \lesssim   2.7$ (we recall that these bounds are derived from a vast collection of current as well as prior experiments and observations). 
At this stage, theoretical spectra can be produced using Eq.~(\ref{eq:ER}-\ref{eq:EZ}), but we still need to determine the range of wavenumbers over which the theory is applicable.

In zonostrophic turbulence, we can assume that the theoretical zonal spectrum is applicable between the Rhines scale $k_{Rh}$ and the viscous dissipative scale denoted by $k_{\nu}$. Conversely, the residual spectrum applies within the range from the Rhines scale ($k_{Rh}$) to the injection scale $k_i^E$.
Outside of these scale ranges, the energy density systematically decreases. The viscous dissipative scale is known to be of millimetric order as in the Earth's ocean. 
This prompts us to assume that $k_\nu \gg k_{Rh}$.
As such, we can estimate the energy of the residual and zonal velocities of the subsurface oceans by 
integrating the theoretical spectra in 
Eq.~(\ref{eq:ER}-\ref{eq:EZ}) between wavenumbers $k_{Rh}$ and $k_i$,
\begin{equation}\label{eq:EzEr}
        E_R^s = \frac{3 C_R P_s^{2/3}}{2 \pi^{2/3}} [L_{Rh}^{2/3}-L_i^{2/3}]  \quad \text{and} \quad E_Z^s = \frac{C_Z \beta^2 L_{Rh}^4}{4 \pi^4},
\end{equation}
where we have assumed that $k_{Rh}=\pi/L_{Rh}$.
The typical magnitudes of the residual and zonal velocities, are obtained using the expressions $U_r = \sqrt{2E_R^s}$ and $U_z = \sqrt{2E_Z^s}$.
The injection scale $L_i$ which enters the residual velocity can be estimated using the critical azimuthal wavenumber at onset \citep{barik23}. Since $L_i \sim E^{1/3} D$ in the limit of $E\ll 1$ relevant to the icy satellites, $L_i$ is expected to be $3$ to $4$ orders of magnitude smaller than the ocean thickness.
Unfortunately, direct measurements of the Rhines scale, indicative of jet sizes, are unattainable due to the unobservable nature of the flow beneath the ice crusts of the satellites.

To tackle this challenge, an upper bound for the jets velocity can be derived from the energetics of the convective flow, i.e. an equilibrium between the injected buoyancy power per unit mass and dissipative processes \citep{jansen23}.
Further assuming that dissipation can be mostly attributed to the drag of the zonal flows near the boundaries $\mathcal{F}_J(U_z)$ yields
\begin{equation}
\mathcal{F}_J(U_z) \sim P_s\,.
\label{eq:balance_drag}
\end{equation}
To account for the latter, \citet{jansen16} assumed a parameterized turbulent quadratic drag of the form $\mathcal{F_J}(U_z) \approx C_D D\,U_z^3$. Using drag coefficients $C_D$ calibrated to model the Earth's ocean, \citet{jansen23} and \citet{kang24} then provide zonal flow velocity estimates ranging from a few mm/s for Enceladus to a few cm/s for Europa.

The applicability of such quadratic boundary drag formulation to the subsurface oceans of the icy satellites remains however unclear. In particular, outside the tangent cylinder,
friction processes mostly involve the drag of quasi-geostrophic zonal jets on the external liquid-ice interface. Since the surface roughness of such ice caps is currently unknown, estimates for the drag coefficients remain speculative.

In light of these considerations, we here rather
assume that dissipation mostly occurs via Ekman friction at the external boundary.
This frictional mechanism can be conceptualized as the dissipation of energy by the Ekman layer with a typical thickness $\epsilon = \sqrt{\nu/\Omega}$ under no-slip boundary conditions.
Based on the formulation of Ekman pumping by \cite{greenspan68} and assuming flow geostrophy in a spherical shell, we derive in \ref{Ap:ZonalVelocity} the following formulation
\begin{equation}
\mathcal{F}_{J}(U_z)=    \dfrac{(r_o \nu \Omega)^{1/2}}{\left(r_o^2-s^2\right)^{3/4}
    } U_z^2,
    \label{eq:Ek_friction}
\end{equation}
which holds for $s \geq r_i$ \citep[see, e.g.][]{gillet06}. This enables us to get an upper bound for the geostrophic zonal flow velocity which does not involve an arbitrary drag coefficient

\begin{equation}\label{eq:MAXU}
    \max (U_z) = \sqrt{P_s \frac{\left(r_o^2-s_\text{mid}^2\right)^{3/4}}{r_o^{1/2}(\nu \Omega)^{1/2}}}.
\end{equation}
The geometric factor $\left(r_o^2-s_\text{mid}^2\right)^{3/4}$ is estimated at mid-radius, as is done for the $\beta$ parameter in Eq.~\eqref{eq:beta}.
It is essential to note that this formulation is only applicable outside the tangent cylinder, where zonal jets primarily develop in our numerical simulations.
Also, the derivation of Eq.~\eqref{eq:MAXU} does not account for possible large-scale topography that would make the upper boundary non-spherical.
Assuming that the energy injected by buoyancy power is transferred upscale through non-linear processes and that Ekman friction serves as the physical mechanism arresting the upscale transfer of energy, Ekman friction defines the largest scale of the system, namely the Rhines scale $L_{Rh}$. In other words, we can solve for the Rhines scale in Eq.~\eqref{eq:EzEr}, considering $\max(U_z)$ from Eq.~\eqref{eq:MAXU} as the upper bound for the zonal velocity.

Similar to velocity estimates, one should regard Rhines scale predictions as upper bounds.  
In addition, solving for Eqs.~\eqref{eq:EzEr} and \eqref{eq:MAXU} does not prevent the Rhines scale from exceeding the ocean depth ($L_{Rh} > D$). 
Such predictions are inherently overestimated, since our derivation 
of the upper bound for the zonal velocity overlook Ekman friction 
inside the tangent cylinder (refer to \ref{Ap:ZonalVelocity}). 
The changes of $h(s)$ across the tangent cylinder prevent determining
a unique relevant value for the
zonal velocity upper bound that would incorporate stresses from both inside and outside the tangent cylinder. 
Hence, in the ongoing analysis, Rhines scale predictions 
must be regarded as possibly overestimated when exceeding the ocean depth.
With this limitation in mind, we present in Table~\ref{tab:predictions} our estimates for each satellite of the upper bound of the geostrophic zonal velocity 
$\max(U_z)$, the associated Rhines scale $L_{Rh}$ and the residual velocity derived from Eq.\eqref{eq:EzEr}.

\begin{table*}
\centering
\caption{Water shell structural properties of the icy satellites along with our velocity predictions. The buoyancy power is estimated using Eq.~\eqref{eq:buoyancy}. 
The ocean depths, denoted as $D$, are derived from the interior model properties provided by \cite{vance18} and references therein. The upper bound of the geostrophic zonal velocity $\max(U_z)$ is estimated using Eq.~\eqref{eq:MAXU}. The Rhines scale results from the combination of Eqs.~\eqref{eq:EzEr} and \eqref{eq:MAXU}. The residual velocity $U_r$ is determined using Eq.~\eqref{eq:EzEr} and the values are averaged while considering the uncertainty associated with the zonal constant $C_Z$. The transitional scale is determined using Eq.~\eqref{eq:lbeta}, taking into account our calculated value of the topographic $\beta$ parameter following Eq.~\eqref{eq:beta}. 
\label{tab:predictions}}
\begin{tabular}{l r r r r}
 \toprule
   & Enceladus & Titan & Europa & Ganymede \\ [0.5ex] 
  \midrule
\multicolumn{5}{c}{Buoyancy power, $P_s$ ($10^{-13}$ m$^2$/s$^{3}$)} \\ [0.5ex]
MgSO$_4$ 10 wt\% & 0.48, 2.63 & 3.11, 16.25, 20.69 & 17.26, 92.17 & 16.22, 30.11 \\ [0.5ex]
Seawater & 0.04, 0.2 & -  & 18.38, 96.72 & -  \\ [0.5ex]
Water & - & 13.54, 25.26, 28.65 & 14.62, 72.47 & 14.8, 28.07, 156.26 \\ [0.5ex]
\midrule
\multicolumn{5}{c}{Ocean thickness, $D$ (km)} \\ [0.5ex]
MgSO$_4$ 10 wt\% & 13, 63 & 91, 333, 403 & 103, 131 & 287, 493 \\ [0.5ex]
Seawater & 12, 55 & -  & 99, 126 & -  \\ [0.5ex]
Water & -  & 130, 369, 420 & 97, 124 & 119, 361, 518 \\ [0.5ex]
\midrule
\multicolumn{5}{c}{Zonal velocity upper bound, $\max(U_z)$ (m.s$^{-1}$)} \\ [0.5ex]
MgSO$_4$ 10 wt\% & 0.011, 0.047 & 0.149, 0.551, 0.666 & 0.237, 0.599 & 0.43, 0.715 \\ [0.5ex]
Seawater & 0.003, 0.013 & -  & 0.241, 0.605 & -  \\ [0.5ex]
Water & -  & 0.355, 0.713, 0.796 & 0.213, 0.521 & 0.297, 0.616, 1.66 \\ [0.5ex]
\midrule
\multicolumn{5}{c}{Residual velocity, $U_r$ (10$^{-3}$ m.s$^{-1}$)} \\ [0.5ex]
MgSO$_4$ 10 wt\% & 1.45, 4.4  & 9.38, 25.16, 29.07 & 13.86, 29.93 & 20.89, 30.37\\ [0.5ex]
Seawater &  0.48, 1.4 & -  & 14.53, 30.36 & -  \\ [0.5ex]
Water & - & 18.72, 31.05, 33.82 & 12.85, 27.12  & 16.18, 27.43, 60.71 \\ [0.5ex]              
\midrule
\multicolumn{5}{c}{Rhines scale, $L_{Rh}/D$, $C_Z=0.3$} \\ [0.5ex]
MgSO$_4$ 10 wt\% & 0.4, 0.4 & 2.1, 2.1, 2.1 & 1.1, 1.6 & 1.4, 1.3 \\ [0.5ex]
Seawater & 0.2, 0.2 & -  & 1.2, 1.7 & -  \\ [0.5ex]
Water & -  & 2.7, 2.3, 2.3 & 1.1, 1.6 & 1.7, 1.5, 2.0 \\ [0.5ex]
\midrule
\multicolumn{5}{c}{Rhines scale, $L_{Rh}/D$, $C_Z=2.7$} \\ [0.5ex]
MgSO$_4$ 10 wt\% & 0.2, 0.2 & 1.2, 1.2, 1.2 & 0.6, 0.9 & 0.8, 0.8 \\ [0.5ex]
Seawater & 0.1, 0.1 & -  & 0.7, 0.9 & -  \\ [0.5ex]
Water & -  & 1.5, 1.3, 1.3 & 0.6, 0.9 & 1.0, 0.8, 1.1 \\ [0.5ex]
\midrule
\multicolumn{5}{c}{Transitional scale, $L_{\beta}/D$, $C_Z=0.3$}  \\ [0.5ex]
MgSO$_4$ 10 wt\% & 0.09, 0.07 & 0.25, 0.21, 0.21 & 0.14, 0.17 & 0.14, 0.13 \\ [0.5ex]
Seawater & 0.05, 0.04 & -  & 0.14, 0.18 & -  \\ [0.5ex]
Water & -  & 0.29, 0.22, 0.22 & 0.14, 0.17 & 0.2, 0.14, 0.18 \\ [0.5ex]
\midrule
\multicolumn{5}{c}{Transitional scale, $L_{\beta}/D$, $C_Z=2.7$}  \\ [0.5ex]
MgSO$_4$ 10 wt\% & 0.05, 0.04 & 0.13, 0.11, 0.11 & 0.07, 0.09 & 0.07, 0.07 \\ [0.5ex]
Seawater & 0.03, 0.02 & -  & 0.07, 0.09 & -  \\ [0.5ex]
Water & -  & 0.15, 0.12, 0.11 & 0.07, 0.09 & 0.1, 0.07, 0.09 \\ [0.5ex]
\bottomrule
\end{tabular}
\end{table*}
 
Ganymede, with an ocean which could be as thick as 518 km, stands out as the most promising candidate for generating strong zonal jets. The upper bound for geostrophic zonal velocity reaches over $1$ m/s, while the associated residual velocity remains at approximately $0.06$ m/s. When the ocean depth is reduced to 119 km, the zonal velocity decreases to $\sim 0.3$ m/s, and the residual velocity drops to around $\sim 0.016$ m/s.
In summary, Ganymede can possibly host strong zonal jets, with zonal velocities falling within the range of $0.3\lesssim U_z \lesssim 1.66$ m/s, encompassing all fluid depth configurations. 
Europa and Titan follow with a zonal velocity range of $0.2\lesssim U_z \lesssim 0.6$ m/s and $0.1\lesssim U_z \lesssim 0.8$ m/s respectively, while Enceladus has a reduced zonal velocity falling within the range of $0.003\lesssim U_z \lesssim 0.05$ m/s. We emphasize that these predictions for the zonal velocity only account for the geostrophic component of the flow.

Our predictions for zonal velocities are lower than those provided by \cite{vance21}, which reach up to $U_z\sim 6$ m/s and $U_r \sim 0.11$ m/s for Ganymede. 
On the other hand, our estimates indicate larger velocities compared to the ranges reported by \cite{bire22}, who suggest values between $10^{-3} \lesssim U_z \lesssim 0.7$ m/s 
and $10^{-2} \lesssim U_r \lesssim 0.5$ m/s. In both cases,  estimations of flow velocities 
were solely based on the assumption that the Rossby numbers from numerical simulations, are equivalent to those in the oceans of icy satellites. This reasoning ignores a rescaling of the physical quantities using either scaling laws for the velocity \citep[e.g.][]{gastine16} or spectral
analyses as done here.
These numerical setups also present different geometries (a spherical shell for \cite{vance21} and a spherical wedge shell for \cite{bire22}) and mechanical boundary conditions that may explain the discrepancies. For comparison, \cite{tyler2008strong} demonstrated that tidal forcing on subsurface oceans triggers large-amplitude Rossby waves with typical velocities ranging from $0.086$ to $0.84$~m/s.

As already mentioned, \cite{jansen23} and \cite{kang24} derived an upper bound for 
the zonal flow velocity using a power balance between the buoyancy power and boundary 
friction. Their parameterized turbulent boundary layer friction 
yields velocity estimates for Enceladus and Europa roughly 5 times lower than ours, a stark contrast to which we shall come back in the discussion below. 

Our velocity predictions are associated with projections of the Rhines scale, 
whose uncertainty is contingent upon the zonal constant ($0.3 \lesssim C_Z \lesssim 2.7$). 
On Europa, the typical jets size falls within the range $0.7D \lesssim L_{Rh} \lesssim 1.4D$ on average. The lower estimate suggests ocean dynamics 
resembling our simulation at $E = 10^{-6}$ and $Ra=2 \times 10^{10}$ 
presented in Fig.~\ref{fig:ZonalMaps}, where the prograde equatorial jet spans more than half of the ocean depth. 
The upper estimate, however, 
suggests a single equatorial jet whose typical size approaches the ocean depth. 
If such a configuration has been reported in numerical models with free-slip boundary conditions \citep{soderlund19}, 
it remains to be seen under no-slip conditions in the presence of Ekman friction. 
With Rhines scale predictions in the range $0.9D \lesssim L_{Rh} \lesssim 1.6D$ on average, Ganymede also appears to be a promising candidate for a single-jet solution outside the tangent cylinder. For Enceladus, however, the typical jets size is reduced to the range $0.15D \lesssim L_{Rh} \lesssim 0.3D$, indicating ocean dynamics reminiscent of our multiple jets simulation at $E = 10^{-6}$ and $Ra=5 \times 10^{9}$, or possibly $Ra=10^9$ (refer to Fig.~\ref{fig:ZonalMaps}). Discussing the size of jets on Titan is more intricate since our predictions 
consistently yield $L_{Rh} > D$.  
Our overall Rhines scale predictions is in line with the regime diagram presented in Fig.~\ref{fig:RavsEk}: smaller Rossby numbers and hence smaller jets in Enceladus reflect the stronger influence of rotation on its ocean dynamics. In contrast, Europa, Ganymede and Titan, straddling the transitional and non-rotational regimes, likely exhibit much larger zonal jets.

In our analysis, Eq.~\eqref{eq:EzEr} also provides an estimate of the residual velocity. Across all satellites, the residual velocity is approximately one order of magnitude lower than the zonal velocity, reaching a magnitude of few mm/s. These estimates arise from integrating the residual spectrum over the range $k_{Rh} \leq k \leq k_{i}^E$. 
In the predictions of the residual velocity, the uncertainty stems from the determination of the Rhines scale, influenced by the uncertainty in $C_Z$, 
which yields standard deviations of the order
$\approx 5 \times 10^{-3}$ m/s for Titan, Europa, and Ganymede, and $\approx 5 \times 10^{-4}$ m/s for Enceladus.

\begin{figure*}[h!]
  \centering
  \includegraphics[width=\textwidth]
    {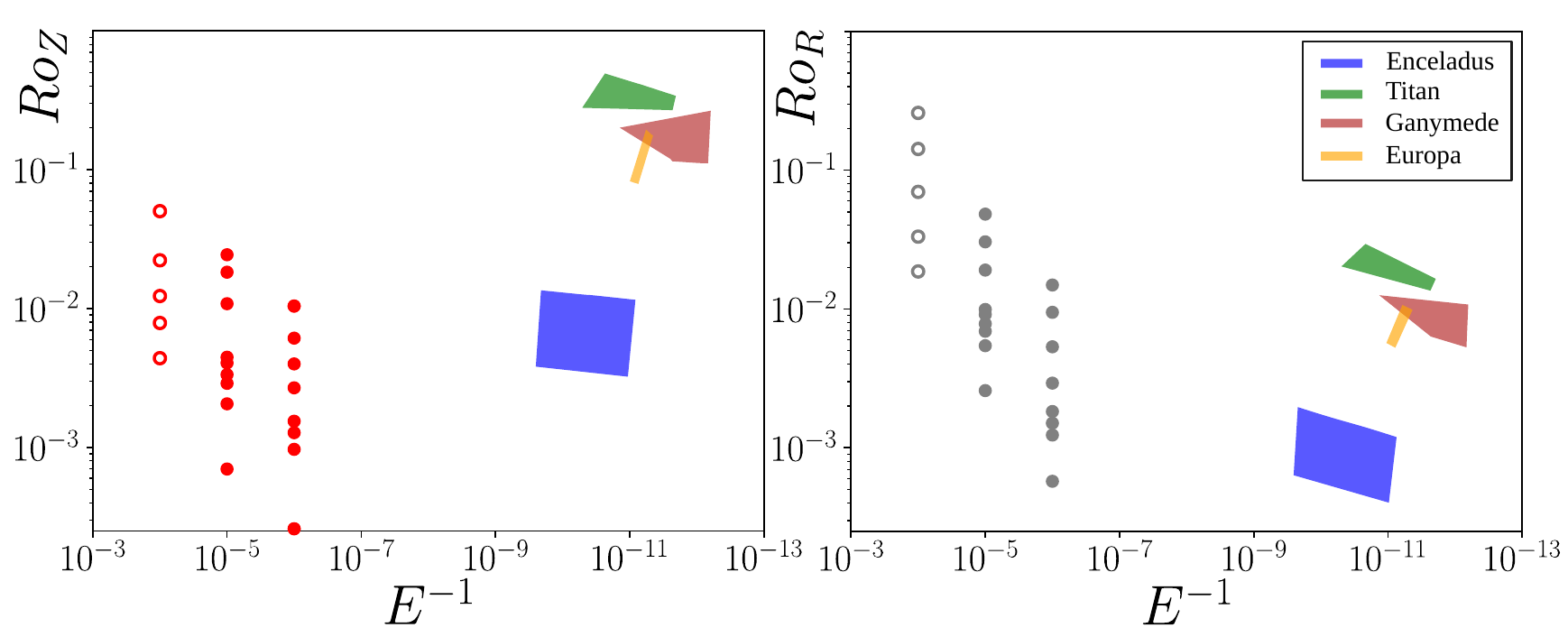}
    \caption{Regime diagrams using Rossby and Ekman numbers from the simulations (dots) and predictions of the subsurface oceans (polygons). The left panel corresponds to the zonal velocity 
    ($Ro_Z$), while the right one corresponds to the residual velocity ($Ro_R$). In each panel, the empty circles indicate values where $R_{\beta}<1$, while the filled-in dots correspond to $R_{\beta}>1$. \label{fig:RovsE} 
    }
\end{figure*}

Figure~\ref{fig:RovsE} provides a comprehensive summary of our numerical simulations and predictive findings. It shows the zonal Rossby number $Ro_Z = U_z/\Omega D$  together with the residual Rossby number $Ro_R = U_r/\Omega D$ computed from our numerical simulations and velocity predictions.  
The size of the colored polygons reflect the uncertainties on internal properties and water compositions.
These diagrams illustrate that in our numerical simulations, a decrease in the Ekman number strengthens the zonal Rossby number at the expense of the residual one. For $E = 10^{-6}$, flows obtained in numerical simulations correspond to $Ro_R\approx Ro_Z$.
Our predictions indicate that this trend continues for the icy satellites, ultimately reaching a state where $Ro_Z \gg Ro_R$.
In such a regime, the turbulence is strongly zonostrophic, and the Rhines scale is much larger than the transitional scale ($L_{Rh} \gg L_{\beta}$, see Table~\ref{tab:predictions}).
In some cases, the theoretical zonostrophy index of the subsurface oceans ($L_{Rh}/L_{\beta}$) can even exceed ten (not reported here). However, it is important to note that this may not entirely reflect reality. As long as we only consider thermal convection, we overlook other sources of turbulence, such as mechanical and electromagnetic forcings, as well as double-diffusive convection \citep{tyler2008strong,vance05,gissinger19}. 
These supplementary sources will enhance the turbulent mixing within the subsurface oceans, likely reinforcing the residual energy of the flow, and consequently, modifying the zonal energy balance through 
cascade effects. Future numerical simulations could incorporate these additional energy 
sources, leading to a revision of the current version of our Rossby-Ekman diagram.

\section{Discussion}\label{sec:Discussion}
We have investigated 21 numerical simulations of rapidly-rotating convection in a spherical shell of radius ratio $r_i/r_o = 0.8$, which is in the low range expected for the subsurface oceans in the Solar System.
Extending previous numerical studies, we have explored a range of Ekman and Rayleigh numbers ($10^{-6} \leq E \leq 10^{-4}$ and $ 10^7 \leq Ra \leq 2 \times 10^{10}$) leading to the formation of multiple zonal jets with no-slip boundary conditions. At the lowest Ekman number, 
the kinetic energy contained in the zonal flow is comparable to the residual energy
(see Fig.~\ref{fig:IntgQuan}). 
This tendency would not have been so evident in earlier studies with Ekman numbers of $\mathcal{O}(10^{-4}-10^{-5})$, if those studies did not assume free-slip boundary conditions.

We took advantage of our new numerical setup to extend
the theory of zonostrophic turbulence to deep-seated convection in a rotating sphere.
We recall that this theory has been developed in purely 2D-flows, thanks to numerical simulations on the sphere \citep{sukoriansky02}. 
Only recently has the theory of zonostrophic turbulence been extended 
to shallow-water simulations \citep{cabanes20a}, mechanically forced laboratory experiments \citep{cabanes17,lemasquerier23}  and atmospheric observations \citep{young17}. However, it is not obvious that this theory would also apply to deep-seated convection in a spherical shell where the $\beta$ parameter substantially changes with the cylindrical radius.

To assess its applicability, we develop a flow decomposition in spectral space using cylindrical harmonic functions, namely the Hankel and Weber-Orr transforms. 
We argue that, given the axial symmetry of the zonal jets about the rotation axis, cylindrical harmonics are better suited for deep models of rapidly-rotating turbulence in spherical geometry.
Thanks to this spectral analysis, we show that both zonal and residual flow components are well accounted for by the theory of zonostrophic turbulence for simulations with $E\leq10^{-5}$ and within the limits of the Rayleigh numbers explored in this study.
For $E\leq 10^{-5}$, present day numerical models cannot effectively reach the non-rotating limit depicted in Fig.~\ref{fig:RavsEk}, which precludes the exploration of the plausible failure of zonostrophic theory in this regime.
We also stress that only the geostrophic component of the zonal energy was found to adhere to the zonostrophic theory. 
These different factors introduce uncertainties regarding the applicability of the zonostrophic theory when trending towards the non-rotating limit of the diagram, a region susceptible to be relevant for several subsurface oceans.
Within these limits, our study demonstrates that it is possible to evaluate the residual and the geostrophic zonal energy of the flow by estimating the available buoyancy power of the convecting fluid, its Ekman number, and an estimate of the $\beta$ parameter.

Assuming that the framework of zonostrophic turbulence can be extrapolated to the more extreme parameters of the icy satellites, we deliver in Table~\ref{tab:predictions} velocity predictions for four subsurface oceans, namely, Enceladus, Titan, Europa, and Ganymede.
Due to the lack of direct constraint on the jets size, necessary for estimating the zonal energy using zonostrophy theory, we define an upper bound for the zonal velocity using energetic constraints
in the same vein as \citet{jansen23}.
This methodology allows us to predict the Rhines scale for each satellite, along with their residual energy.

We demonstrate that Europa and Ganymede may exhibit zonal jets with a typical size approaching the ocean depth, while Enceladus is likely characterized by multiple jets located outside the tangent cylinder.
Predictions for Titan present additional challenges, as our jet size estimates consistently exceed the ocean depth. This poses a complication, given that our theoretical approach is built upon the assumption of 
zonal jets 
confined outside the tangent cylinder
and given that there is no evidence of numerical models with rigid boundaries developing jets broader than the shell thickness.
Furthermore, our analysis reveals that for subsurface oceans, the upper bound 
for the geostrophic zonal velocities ranges from $\sim 0.1$ to $1.6$~m/s, while the residual velocities do not exceed $\sim~0.003$ to $0.06$~m/s.
The exception may be Enceladus, the smallest of the four satellites, 
whose zonal velocity ranges from $\sim 0.003$ to $0.05$~m/s and 
residual velocity ranges from $\sim 0.0005$ to $0.004$~m/s. 
We emphasize that our estimates of the zonal velocity rests entirely on the geostrophic component of the flow. 
Ageostrophic contributions can further enhance our 
estimates, prompting the need for additional investigations in that regard.

Our theoretical approach places a primary focus on the kinetic energy budget within convective flows. It is similar to
the approach of \citet{jansen23}, while diverging from earlier analyses that simply 
assumed a direct equivalence between the Rossby number in numerical simulations and that in real oceanic flows \citep[e.g.][]{amit20, soderlund19,bire22}.
The difference between our study and those by \citet{jansen23} and \citet{kang24} stands in the 
treatment of the boundary drag: while they retain a quadratic turbulent 
drag hypothesis initially tailored to model Earth's ocean, we rather assume that boundary 
friction is governed by Ekman pumping outside the tangent cylinder of the subsurface oceans. At this stage of our knowledge of the ice-water interface in these hidden ocean worlds, it is however fair to say 
that no definite argument can be provided in favor of one hypothesis
or the other. This leaves us with a possible factor of $3-5$ between both velocity upper bounds, a prospect for future investigations.

Additionally, it is worth noting that prior numerical simulations employing free-slip boundary conditions overlook Ekman friction, a process
we underscore as crucial for stabilizing the jets. Without considering this frictional effect, zonal jets tend to equilibrate with viscous dissipation in the fluid bulk (refer to Eq.~\eqref{eq:upqg} in the Appendix) and typically reach system scale \citep[e.g.][]{soderlund19}.
If we extrapolate to the Ekman numbers relevant to icy satellites, where 
fluid viscosity is significantly lower, we encounter unrealistic scenarios where the residual energy would become negligible compared to the zonal energy. This speculation is influenced by the findings of \cite{soderlund19}, who already demonstrated
at $E=3\times 10^{-4}$
a partition of $E_Z \sim 100 E_R$, this factor of $100$ being already comparable to our predictions for the sub-surface oceans. 
Therefore, we advocate for the necessity of employing no-slip boundary conditions in simulating subsurface ocean dynamics.

One of the main outcome of this study is to provide a Rossby-Ekman diagram that emphasizes flow velocities independently of the nature of the energetic sources that drive the flow. This contrasts with the Rayleigh-Ekman diagram introduced by \cite{gastine16}. Currently, this Rossby-Ekman diagram allows us to locate our numerical simulations alongside with the predictions for subsurface oceans. We identify a continuous regime transition along the Ekman axis, shifting from a non-zonostrophic or quasi-3D turbulent regime at $E=10^{-4}$ to a zonostrophic turbulent regime at $E=10^{-5}$ and $10^{-6}$. The transition threshold occurs when the zonostrophy index equals one, as illustrated in Fig.~\ref{fig:EREZ} with the empy and filled-in circles. However, no such transition is observed along the Rossby axis, as the zonostrophy index remains above unity for all simulations at $E=10^{-5}$ and $10^{-6}$. This is due to the high computational cost of 3D turbulent experiments, which prevents us from further exploring low Ekman and high Rossby numbers $Ro \sim \mathcal{O}(1)$, a regime possibly relevant for icy satellites such as Titan, Ganymede, and Europa.
Within this limit, the global fraction of non-geostrophic energy increases,  prompting questions about the continued validity of zonostrophic theory. Further investigations are required to elucidate these uncertainties. 

Consequently, we advocate further exploration of the regime diagrams presented in Figs.~\ref{fig:RavsEk} and \ref{fig:RovsE} through deep-seated turbulent experiments. This effort should aim to delineate the dynamical boundaries of the zonostrophic regime of turbulence. For this purpose, laboratory experiments offer the potential for more turbulent flows, elevating the current Rossby number, while quasi-geostrophic simulations can reduce the computational cost to explore lower Ekman numbers \citep[see][]{barrois22,lemasquerier23}.

In conclusion, it is crucial to stress that the estimate of an upper bound for the zonal flow velocity is instrumental to bring some predictive power to the theory of zonostrophic turbulence.
This is required by the absence of direct observational data regarding the size of ocean jets, or equivalently, the Rhines scale. 
To address this limitation, there is a possibility that direct measurements of surface heat transfer or ice topography may carry indications of the underlying jet flows and, by extension, their latitudinal size and location. This approach has been previously investigated in studies conducted by \cite{soderlund14,kvorka18,amit20,kvorka22} and \cite{terra23}. Such measurements are planned for Europa and Ganymede, as they are the main targets of ESA's JUICE mission and NASA's Europa Clipper mission. 
Should these future missions yield information about the typical size of Europa's and Ganymede's ocean jets, they would 
offer additional invaluable constraints to enhance the accuracy of our theoretical predictions regarding flow velocities.

\section*{Acknowledgments}
The authors thank Wanying Kang and an anonymous reviewer for their suggestions that helped improve the manuscript.
The authors acknowledges the support of the French Agence Nationale de la Recherche (ANR), under grant ANR-19-CE31-0019 (project RevEarth). Numerical computations were performed on the S-CAPAD/DANTE platform at IPGP.
This work has been funded by ESA in the framework of EO Science for Society, through contract 4000127193/19/NL/IA (SWARM + 4D Deep Earth: Core).

\appendix
\section{A bound for the axisymmetric azimuthal velocity in the rapidly-rotating regime}\label{Ap:ZonalVelocity}
In the rapidly-rotating, or quasi-geostrophic, regime, the $s$ and $\phi$ components
of the quasi-geostrophic velocity $\mathbf{u}^g$ are invariant in the direction
of global rotation $z$.  An equation for the time-evolution of the
axisymmetric azimuthal velocity is obtained by averaging the Navier-Stokes
equation
\begin{equation}
  \rho \frac{\partial \mathbf{u}}{\partial t} + \rho (\mathbf{u} \cdot \nabla ) \mathbf{u} + 2\rho \Omega \mathbf{e}_z\times \mathbf{u}  =    - \nabla P + \rho \nu \nabla^2 \mathbf{u} + \rho \alpha T \mathbf{g}.
  \label{eq:NS}
\end{equation}
over a so-called geostrophic cylinder $\mathcal{C}(s)$, 
whose cylindrical radius $s$ is comprised between $r_i$ and $r_o$. 
In the $z$ direction, $\mathcal{C}(s)$ extends between $z=-h^++\epsilon$ and $z=h^+-\epsilon$, where $h^{+}=\sqrt{r_o^2 -s^2}$ and $\epsilon$ is the thickness of the viscous Ekman boundary layers that develop over 
the no-slip outer spherical boundary. Denoting the average over $\mathcal{C}(s)$ by $\langle \cdot \rangle_{\mathcal{C}(s)}$, we obtain 

\begin{equation}
\begin{aligned}
\left[ 
\frac{\partial}{\partial t}
+
 (\nu \Omega)^{1/2} \frac{r_o^{1/2}}{h^{+^{3/2}}} 
-\nu
\left(
\frac{1}{s^2} 
\frac{\partial}{\partial s}
s^3
\frac{\partial}{\partial s}
\frac{1}{s}
    \right)
\right] \langle u_{\phi}^{g} \rangle_{\mathcal{C}(s)}
= \\
-\frac{1}{s^2}\left\langle  \partial_s\left( s^2u_s^{g}  \up^{g} \right)  \right\rangle_{\mathcal{C}(s)}
\end{aligned}
    \label{eq:upqg}
\end{equation}
A secondary flow, driven by the so-called Ekman
pumping, occurs within Ekman boundary layers and permeates the bulk of the fluid \citep{greenspan68}. Ekman pumping is responsible
for the second term on the left-hand side of Eq.~\eqref{eq:upqg}, via the calculation of $ \langle u_s^{g} \rangle_{\mathcal{C}(s)}$, see e.g.
\cite{schaeffer2005} and \cite{gillet06} for more details. The term on the right hand-side of Eq.~\eqref{eq:upqg} is the nonlinear
Reynolds stress term, that will be denoted by $\mathcal{R}$ in the following.
The balance for the axisymmetric zonal energy is obtained by multiplying Eq.~\eqref{eq:upqg} by $\langle \up^{g} \rangle_{\mathcal{C}(s)}$,
\begin{eqnarray*}
\frac{\partial}{\partial t}\left( \frac{1}{2} \langle \up^{g} \rangle_{\mathcal{C}(s)}^2\right)  &=&
\mathcal{R}  \langle \up^{g} \rangle_{\mathcal{C}(s)} - (\nu \Omega)^{1/2} \frac{r_o^{1/2}}{h^{+^{3/2}}} 
\langle \up^{g} \rangle_{\mathcal{C}(s)}^2
\\
&+&\nu
    \left[
    \frac{1}{s^2} 
\frac{\partial}{\partial s}
s^3
\frac{\partial}{\partial s}
\left(
\frac{\langle u_{\phi}^{g}\rangle_{\mathcal{C}(s)}}{s}
    \right)  \right] \langle \up^{g} 
    \rangle_{\mathcal{C}(s)} .
\end{eqnarray*}
The axisymmetric zonal energy can increase in response to the first term on the right-hand side, and gets dissipated either by Ekman friction
arising from the Coriolis term or by viscous stresses. 
We shall assume that the former is more effective at dissipating energy at the
scale of jets than the latter. On time average, this implies that
\begin{equation}
\left| \overline{\mathcal{R} \langle \up^{g} \rangle_{\mathcal{C}(s)} }\right| \approx (\nu \Omega)^{1/2} \frac{r_o^{1/2}}{h^{+^{3/2}}} 
\overline{\langle  \up^{g}  \rangle_{\mathcal{C}(s)}^2}.
\end{equation}
Assuming that all energy injected by the mean buoyancy power $P_s = \alpha g q/\rho C_p$ is channeled into the zonal jets through
nonlinear processes,  namely that $\left| \mathcal{R} \langle \up^{g} 
\rangle_{\mathcal{C}(s)} \right| = P_s  $,
we derive an upper bound for the axisymmetric quasi-geostrophic zonal velocity,
\begin{equation}
   \overline{\langle  u_{\phi}^{g} \rangle_{\mathcal{C}(s)} } \approx \sqrt{P_s \frac{\left(r_o^2-s^2\right)^{3/4}}{r_o^{1/2}(\nu \Omega)^{1/2}}}.
\end{equation}
In the main text, this upper bound of the zonal velocity is denoted by $\max (U_z)$.
The geometric factor $\left(r_o^2-s^2\right)^{3/4}$ is estimated at mid-radius, as is done for the $\beta$ parameter in Eq.~\eqref{eq:beta}.

\section{Properties of the subsurface oceans}

\label{Ap:param}
All the relevant properties of the icy satellites are reported in Table~\ref{tab:Annexe1}. Building upon the methodology outlined by \cite{soderlund19} we can anticipate the convective regime of the icy satellites by estimating their Ekman, Rayleigh, and Prandtl numbers. The Prandtl number, which solely depends on the properties of the fluid, is approximated to be around $Pr \sim 10$ for the oceans of the satellite \citep{abramson01,nayar16}. Calculating the Ekman number is also relatively straightforward, as it only depends on fluid viscosity, the rotation rate, and ocean depth \citep[here the interior model from][is employed]{vance18}. Following on from \cite{soderlund19}, the Rayleigh number requires the knowledge of the superadiabatic temperature contrast $\Delta T$. \cite{soderlund19}, suggest to solve for $\Delta T$ algebraically considering both non-rotating and rapidly-rotating scaling laws for the heat flux $q$. The temperature contrast considering a non-rotating regime is,
\begin{equation}
    \Delta T = 7.3\left( \frac{\nu}{\alpha g \rho C_p} \right)^{1/4} q^{3/4}\,,
\end{equation}
and for the rotating regime,
\begin{equation}
    \Delta T = 2.1\left( \frac{\Omega^4 \kappa}{\rho^2 C_p^2 \nu \alpha^3 g^3} \right)^{1/5} (q^2 D)^{1/5}\,,
\end{equation}
see also \cite{gastine16} for more details. Note that the range of values for the dimensionless quantities given in Table~\ref{tab:Annexe1} correspond
to the estimated minima and maxima.
\begin{table*}
\centering
   \caption{Summary of the properties of the subsurface oceans of the Jovian and Saturnian satellites \citep[see][and references therein for further details]{vance18,soderlund19} \label{tab:Annexe1}}
\begin{tabular}{l r r r r}
 \toprule
   & Enceladus & Titan & Europa & Ganymede \\ [0.5ex] 
  \midrule
$g$ (m/s$^2$) & 0.1 & 1.4 & 1.3 & 1.4 \\ [0.5ex] 
$\Omega$ (s$^{-1}$) & 5.3 $\times 10^{-5}$ & 4.6 $\times 10^{-6}$ & 2.1 $\times 10^{-5}$ & 1.0 $\times 10^{-5}$ \\ [0.5ex]
$\nu$ (m$^2$/s) & $1.8 \times 10^{-6}$ & $1.8 \times 10^{-6}$ & $1.8 \times 10^{-6}$ & $1.8 \times 10^{-6}$  \\ [0.5ex]
$\kappa$ (m$^2$/s) &$1.4 \times 10^{-7}$& $1.8 \times 10^{-7}$ & $1.6 \times 10^{-7}$ &$1.8 \times 10^{-7}$ \\ [0.5ex]
$R$ (km) & 252 & 2575 & 1561 & 2631  \\ [0.5ex]
\midrule
 \multicolumn{5}{c}{Ice Ih thickness, $D_{I}$ (km)} \\ [0.5ex]
 MgSO$_4$ 10 wt\% & 50, 10 & 149, 86, 58 & 30, 5 & 157, 95, 26 \\ [0.5ex]  
 Seawater & 50, 10 & - & 30, 5 & - \\ [0.5ex]
 Water  & 51, 10 & 141, 74, 50 & 30, 5 & 134, 70, 5 \\ [0.5ex] 
 \midrule
 \multicolumn{5}{c}{Ocean thickness, $D$ (km)} \\ [0.5ex]
 MgSO$_4$ 10 wt\% &  13, 63 & 91, 333, 403 & 103, 131 & 24, 287, 493 \\ [0.5ex]
 Seawater & 12, 55 & - & 99, 126 & - \\ [0.5ex] 
 Water  & 11, 53 & 130, 369, 420 & 97, 124 & 119, 361, 518 \\ [0.5ex] 
 \midrule
 \multicolumn{5}{c}{Heat flux, $q$ (mW/m$^2$)} \\ [0.5ex]
 MgSO$_4$ 10 wt\% & 16, 83 & 14, 17, 19 & 24, 123 & 15, 18, 25 \\ [0.5ex]
 Seawater & 16, 82 & - & 23, 121 & -  \\ [0.5ex] 
 Water & 16, 81 & 14, 18, 20 & 24, 119 & 16, 20, 107 \\ [0.5ex] 
 \midrule
 \multicolumn{5}{c}{Density, $\rho$ ($10^3$ kg/m$^3$)} \\ [.5ex]
 MgSO$_4$ 10 wt\% & 1.11, 1.11 & 1.20, 1.23, 1.24 & 1.15, 1.14 & 1.19, 1.23, 1.24 \\ [0.5ex]
 Seawater & 1.02, 1.02 & - & 1.07, 1.07 & - \\ [0.5ex] 
 Water & 1.00, 1.00 & 1.11, 1.14, 1.14 & 1.04, 1.04 & 1.11, 1.14, 1.14 \\ [0.5ex] 
 \midrule
 \multicolumn{5}{c}{Heat capacity, $C_p$ ($10^3$ J/kg/K)} \\ [0.5ex]
 MgSO$_4$ 10 wt\% & 3.6, 3.7 & 2.1, 2.5, 2.8 & 3.3, 3.5 & 2.1, 2.4, 3.0 \\ [0.5ex]
 Seawater & 4.0, 4.0 & - & 3.8, 3.8 & - \\ [0.5ex] 
 Water & 4.2, 4.2 & 3.0, 3.5, 3.6 & 3.9, 3.9 & 3.0, 3.5, 3.7 \\ [0.5ex] 
 \midrule
 \multicolumn{5}{c}{Thermal expansivity, $\alpha$ ($10^{-4}$ K$^{-1}$)} \\ [.5ex]
 MgSO$_4$ 10 wt\% & 1.2, 1.3 & 0.4, 2.1, 2.7 & 2.1, 2.3 & -0.1, 1.9, 3.2 \\ [0.5ex]
 Seawater & 0.1, 0.1 & - & 2.5, 2.5 & - \\ [0.5ex] 
 Water & -0.5, -0.5 & 2.3, 4.0, 4.2 & 1.9, 1.9 & 2.2, 4.0, 4.4  \\ [0.5ex]
 \midrule
 \multicolumn{5}{c}{$\beta$ parameter, $\beta$ ($10^{-10}$ m$^{-1}$.s$^{-1}$)}  \\ [0.5ex]
MgSO$_4$ 10 wt\% & 80.21, 15.65 & 1.0, 0.27, 0.22 & 4.01, 3.14 & 0.68, 0.39 \\ [0.5ex]
Seawater & 87.0, 18.11 & -  & 4.17, 3.26 & -  \\ [0.5ex]
Water & -  & 0.7, 0.24, 0.21 & 4.26, 3.32 & 1.66, 0.53, 0.37 \\ [0.5ex]
\midrule
 \multicolumn{5}{c}{Dimensionless numbers} \\ [0.5ex]
$Pr = \frac{\nu}{\kappa}$ &13&10&11&10\\ [0.5ex]
$E = \frac{\nu}{\Omega D^2}$ & $10^{-12} - 10^{-10}$ & $10^{-12} - 10^{-11}$ & $[5- 9] \times 10^{-12}$ & $10^{-13} - 10^{-11}$\\ [0.5ex]
$Ra = \frac{\alpha g D^3 \Delta T}{\nu \kappa}$ &$10^{16} - 10^{20}$ &$10^{19} - 10^{24}$ &$10^{23} - 10^{24}$ & $10^{22} - 10^{24}$\\ [0.5ex]
$\eta = \frac{(R-D_I-D)}{(R-D_I)}$ &0.74-0.94&0.83-0.96&0.91-0.94&0.8-0.95\\ [0.5ex]
  \bottomrule
\end{tabular}
\end{table*}

\bibliographystyle{elsarticle-harv}


\end{document}